%\input amstex

%\documentstyle{amsppt}
%\magnification=\magstep1
%\baselineskip= 12pt
%\hsize=13cm    
%\tolerance=20000
\tolerance=3000
\magnification=1200
\documentstyle{amsppt}
\catcode`\@=11
\def\logo@{\relax}
\catcode`\@=\active
\loadbold
%\Article
\def\mes{\operatorname{mes}}

\def\arg{\operatorname{arg}}
\def\im{\operatorname{im}}
\def\re{\operatorname{re}}

\def\a{\alpha}

\def\g{\gamma}
\def\b{\beta}
\def\s{\sigma}
\def\k{\kappa}
\def\l{\lambda}
\def\R{{\Cal R}}
\def\A{{\Cal A}}
\def\B{{\Cal B}}
\def\C{{\Cal C}}
\def\D{{\Cal D}}
%\article
\document

\topmatter
\centerline
{\bf STARK-WANNIER TYPE OPERATORS }
\centerline
 {\bf WITH PURELY
SINGULAR SPECTRUM}

%%\endtitle
%\author
%\endauthor
\endtopmatter
\head
Galina Perelman
\endhead
\centerline{Centre de Math\'ematiques}
\centerline{Ecole Polytechnique}
\centerline{F-91128 Palaiseau Cedex}
\centerline{France}

\vskip16pt

\TagsOnRight
\subsubhead Abstract
\endsubsubhead
We consider the one-dimensional Stark-Wannier type
operators
 $$ H=-{d^2\over dx^2}-Fx-q(x)+v(x),\quad F>0,$$
where $q$ is a smooth function slowly growing
at infinity and $v$ is periodic,  $v\in L_1({\Bbb T})$,
with the  Fourier coefficients of the form
$(\ln|n|)^{-\b}$, $0<\b<{1\over 2}$, as $n\rightarrow \infty$.
We show that  
for suitable $q$ 
and $F$ the spectrum of the corresponding operator is purely
singular continuous. This proves 
the sharpness of the a.c. spectrum stability
result obtained in [16].

\rm
\head
1. Introduction
\endhead 
In this paper we consider the following operators on ${\Bbb R}$
$$ H=-{d^2\over dx^2}-Fx -q(x)+v(x),\quad F>0,
\tag 1.1$$
where 
$q$ is a smooth function slowly growing
at infinity:
$$|q^{(k)}(x)|\leq c<x>^{\a-k},\quad \a<1,\quad k=0,1,2,$$
$<x>=(1+x^2)^{1/2}$, and $v$ is periodic, $v(x+1)=v(x)$, 
$v\in L_1({\Bbb T})$,
$$\int_0^1dx v(x)=0.\tag 1.2$$

\par The spectral properties of  this model 
have been extensively discussed
in both mathematical and physical literature,
see 
[1-5, 7, 10, 15-17] and references therein.
If $v=0$
the spectrum of $H$
is purely absolutely continuous:
$$ \sigma(H)=\sigma_{ac}(H)={\Bbb R}.\tag 1.3$$
It is known that 
the spectrum of (1.1)
remains absolutely continuous and  covers the whole axis 
under rather weak smoothness requirements on the potential $v$. 
In fact, (1.3) can be proved
for $v\in H^s({\Bbb T}),\,\, s>0$, see [7, 17]. 
On the other hand there are examples of very singular
periodic perturbations such as an array of $\delta^\prime$ potentials
for which the spectrum has no a.c. component or even is pure point, see
[2,3, 15].
It was argued by Ao [1] that the spectral nature depends on the gap 
structure of the periodic perturbation. He conjectured that 
if the size of the $n$-th gap behaves as $n^{-\a}$
 the spectrum is pure point for $\a<0$
(at least for ``non-resonant'' $F$) and continuous for $\a>0$.
In the critical case $\a=0$, which corresponds to a comb of $\delta$
function potentials, a transition from pure point to 
continuous spectrum is expected as $F$ grows (see also [9, 14]
for related phenomena in the random setting). Furthermore,
the spectral nature in this critical case seems to depend
also on the number theoretical properties of $F$ [7].
In [16] the following sufficient condition for 
the stability of the a.c. spectrum was established.

\proclaim{Theorem 1.1} 
Suppose that 
$$v\in L_1({\Bbb T})\cap H^{-1/2}({\Bbb T}).\tag 1.4$$
Then, an essential support of the absolutely
continuous spectrum of the operator $H$
coincides with the whole axis.
\endproclaim
\par Condition  (1.4) corresponds to $\a>0$ in Ao language.
Some intermediate results were obtained in [17].
\par {\it Remark}. If $v\in L_1({\Bbb T})$ then the essential spectrum
of $H$, $\s_{ess}(H)$, fills up the real axis:
$$\s_{ess}(H)={\Bbb R}.\tag 1.5$$
The proof of this fact is a simple compactness type
 argument,
for the sake of completeness we outline it in
appendix 3.

\par The goal of the present paper is to show that the result of 
theorem 1.1
is optimal.
We consider  operator (1.1) with $q(x)=\k\ln<x>$:
$$H=-{d^2\over dx^2}-Fx -\k\ln<x>+v(x),\quad F>0.
\tag 1.6$$
To avoid cumbersome general conditions we consider the case
where
  the Fourier coefficients of
$v$, $\hat v(n)=\int_0^1 dx e^{i2\pi nx}v(x)$, satisfy for some 
$n_0\geq 2$
$$\hat v(n)=v_0(\ln n)^{-\beta},\quad n\geq n_0,\,\,\, v_0\neq 0,\,\,\,
0<\beta<{1\over 2}.\tag 1.7$$
Remark that (1.7) implies
$$v\in C^\infty({\Bbb R}\setminus {\Bbb Z})\cap L_{1,loc}({\Bbb R}).$$

 Our main result is the following theorem.
\proclaim{Theorem 1.2}
For ${\pi ^2\over F}\in{\Bbb Q}$   
and $\k\neq 0$ the spectrum of 
{\rm (1.6)} is purely singular continuous.
\endproclaim
\par {\it Remarks}. 
\par 1. If ${\pi ^2\over F}$
is rational and $\k= 0$ the spectrum of (1.6) 
 is  absolutely continuous
except, may be some discrete set of the eigenvalues, 
see [7] and also subsection 2.1.
\par 2. With some extra efforts the methods of the present paper can 
be made work also for $\b={1\over 2}$. 
\par The rest of the paper is devoted to the proof of theorem 1.2. 
The basis for our analysis is the asymptotic constructions of [7]
that we combine with the ideas of the works [6, 8, 14].
In [7] operator (1.1) with $q=0$ and 
$$v(x)=V\sum\limits_l \delta(x-l)$$
was considered. It was shown that the spectral properties of
this model can be characterized by the asymptotic behavior
as $l\rightarrow +\infty$
of the solutions of the following discrete system
$$\psi(l+1)=W(l)\psi(l),\quad \psi(l)\in {\Bbb C}^2,\tag 1.8$$
where
$$W(l)=e^{i\Gamma(l)\s_3}S(l)e^{-i\Gamma(l)\s_3},
\quad \det S(l)=1,\tag 1.9$$
$$\Gamma(l)={(\pi l)^3\over 3F} +{\pi l (E-V)\over F},
\quad S(l)=\pmatrix 1+dl^{-1}& rl^{-1/2}\cr
\bar rl^{-1/2}&1+dl^{-1}\cr\endpmatrix,$$
$d\in{\Bbb R}$, $r\in{\Bbb C}$.
\par 
When applied to (1.6) (with  ${\pi ^2\over F}\in {\Bbb Q}$ and $\k\neq 0$)
the constructions of [7]
lead to   model (1.8), (1.9) with
$$ \Gamma (l)\sim \rho(E)\Lambda^l,\quad S(l)=I+O(l^{-\beta}),
\quad \Lambda>1.$$
This system  is close to the case of discrete Schr\"odinger operators with
strongly mixing decaying potentials and the  techniques of [6, 8, 14]
can be applied. 
\head 
2. Reduction to a strongly mixing model.
\endhead
\subhead
2.1. Some preliminary reductions
\endsubhead
\subsubhead
2.1.1. Pr\"ufer type coordinates
\endsubsubhead

To derive the desired spectral properties we
study the solutions of the equation
$$-\psi^{\prime\prime}-Fx\psi -q(x)\psi +v(x)\psi=E\psi,
\quad E\in {\Bbb R},\tag 2.1$$
the link between the behavior of solutions and spectral results
being provided by Gilbert-Pearson subordinacy theory [12, 13].
As soon as the a.c. spectrum is concerned the whole line operator
can be replaced by  a right half -line operator with some self-adjoint
boundary conditions in a point $x=R$. Indeed, since 
$-{d^2\over dx^2}+v$ is bounded from below
and $-Fx-q \rightarrow +\infty$ as $x\rightarrow -\infty$,
the  spectrum of the left half-line operator 
is discrete. So, the absolutely continuous parts of the whole line operator
and the right half-line operator are unitarily equivalent
and by subordinacy theory 
to prove theorem 1.2 it is sufficient to show that
\par (i)
for a.e. $E\in {\Bbb R}$ there exist a solution  
subordinate on the right;
\par (ii) for any  $E\in {\Bbb R}$
there is no solution which is in $L_2$ on the right.
\newline 
Recall that a real solution $\psi_1(x,E)$
of (2.1) is called subordinate on the right if for any other
linearly independent solution $\psi(x,E)$ one has
$$\lim\limits_{N\rightarrow +\infty}
{\int_0^N dx |\psi_1(x,E)|^2\over  \int_0^N dx |\psi(x,E)|^2}=0.$$
The main part of the paper is devoted to the proof of part (i),
 (ii) being obtained as a simple by-product of our constructions,
see proposition 2.1.
\par To study the asymptotic behavior of the solutions of (2.1) we 
employ Pr\"ufer type transformation  which is known to be extremely 
useful in the cases of small or decaying randomness, see [6, 8, 11, 14].
\par 
First, we perform a Liouville transformation in (2.1), setting
for $x$ sufficiently large
$$ \psi(x,E)=p^{-1/2}Q(\xi(x)),
 \quad p=(Fx+q+E)^{1/2},\quad \xi^\prime(x)=p(x,E), $$
$$\xi={2\over 3F}(Fx+q+E)^{3/2}
-{2\k\over F}(Fx+q+E)^{1/2}+O(x^{-1/2}\ln x),\quad x\rightarrow +\infty.$$
The resulting function $Q$  satisfies the Schr\"odinger equation
$$-Q_{\xi\xi}-Q+{V\over p}Q=0,\tag 2.2$$
$$V=vp^{-1}+{1\over 4}q^{\prime\prime}p^{-3}-
{5\over 16}(F+q^\prime)^2p^{-5}.$$
Let us further apply a Pr\"ufer type transformation:
$$Q=R\sin\theta,$$
$$Q_\xi=R\cos \theta.$$
Then $R$, $\theta$ satisfy
$${d\over dx}\ln R={1\over 2}V\sin 2\theta ,$$
$${d\over dx}\theta=p-V\sin^2\theta .\tag 2.3$$
It is not difficult to check that
$$\int_{N_1}^{N_2} dx \psi^2\leq \int_{N_1}^{N_2} dx p^{-1}R^2
\leq C\int_{N_1}^{N_2}dx \psi^2,\tag 2.4$$
provided $N_2\geq  N_1$
are sufficiently large:
$N_1\geq C$. 
\par 
The constants $C$ here and below are uniform with respect to
$E$ in compact subsets of ${\Bbb R}$ but may depend on $q$ and $v$.

\subsubhead
2.1.2. Reduction to a discreet system
\endsubsubhead 
First we are going to analyse $R$ along a sequence $x=x_l$,
where $x_l=[\tilde x_l]-1/2$, $\tilde x_l$ being  defined by the relation
$$F\tilde x_l+q(\tilde x_l)+E=\pi^2\left (l-{1\over 2}\right)^2.$$
We start by a sequences of auxiliary estimates.
\proclaim{Lemma 2.1}
For $x\in [x_l,x_{l+1}]$, $v\in L_1({\Bbb T})$
and any $0<\nu<{1\over 2}$ one has the following inequalities
$$|\int_{x_l}^x dy\chi_l(y)e^{2i\xi(y)}v(y)|\leq c(l^{1/2}|\hat v_l|+
\|v\|_{ L_1({\Bbb T})}),$$
$$|\int_{x_l}^x dye^{2i\xi(y)}v(y)(1-\chi_l(y))|\leq c(l^{\nu}|\hat v_l|
+\|v\|_{ L_1({\Bbb T})}).$$
Here $\chi_l(x)=\chi(l^{-1+\nu}(x-X_l))$, 
$\chi\in C_0^\infty$, $\chi(s)=\chi(-s)$,
$$\chi(s)=\cases 1, \quad \roman{if}\,\,\, |s|\leq 1,\cr
0\quad \roman{if}\,\,\, |s|\geq 2,\cr   \endcases$$
$X_l$ solves the equation
$$FX_l+q(X_l)+E=\pi^2l^2.$$
\endproclaim
 See appendix 1 for the proof.

\par It follows immediately from lemma 2.1 and (2.2) that
for $x\in [x_l,x_{l+1}]$, and any $ g\in L_1({\Bbb T})$
$$\int_{x_l}^x dye^{2i\theta(y)}g(y)=
e^{2i\varphi(x_l)}\int\limits_{x_l}^x dy e^{2i\xi(y)} g(y) $$
$$
-2i\int\limits_{x_l}^x dy e^{2i\varphi(y)}V(y) \sin^2\theta(y)
\int_{y}^xds
e^{2i\xi(s)} g(s)=O(l^{1/2})\|g\|_{ L_1({\Bbb T})},
\quad \varphi=\theta-\xi.\tag 2.5$$
Here and below the constants in $O(\cdot)$ 
are independent of
the choice of initial conditions in (2.1),  uniform with respect to 
$E$ in compact subsets 
 of ${\Bbb R}$, but may depend on $q$ and $v$.

\par Consider the expression $\ln {R(x)\over  R(x_l)}$. 
By (2.2), (2.3) and lemma 2.1,
$$\ln{ R(x)\over R(x_l)}=
{1\over 2}\im\left(\int\limits_{x_l}^x dy  e^{2i\theta} vp^{-1}\right)
+O(l^{-4})={1\over 2\pi l}\im
\left(e^{2i\varphi(x_l)}\int\limits_{x_l}^x dy  
e^{2i\xi} v\right)$$
$$-
{1\over (\pi l)^2}
\re\left(
\int\limits_{x_l}^x dy e^{2i\varphi(y)}v(y) \sin^2\theta(y)
\int_{y}^xds
e^{2i\xi(s)} v(s)\right)+O(l^{-3/2}),\tag 2.6$$
which means in particular that
$$\ln {R(x)\over  R(x_l)}=O(l^{-1/2}).\tag 2.7$$
Therefore,
 if we control $R(x_l,E)$ we will have
sufficient control for all $x$. 
\par It follows from lemma 2.1 and (1.2), (2.5) that
$$\int\limits_{x_l}^x dy e^{2i\varphi(y)}v(y)\int^{x}_yds
e^{2i\xi(s)} v(s)$$
$$=\int\limits_{x_l}^x dy e^{2i\theta(y)}v(y)
\int\limits_{x_l}^yds
 v(s)+O(l^{1/2})=O(l^{1/2}),\tag 2.8$$
$$\re 
\int\limits_{x_l}^x dy e^{-2i\xi(y)}v(y)\int^{x}_yds
e^{2i\xi(s)} v(s)={1\over 2}\left|\int_{x_l}^xdy
e^{2i\xi(y)} v(y)\right|^2,$$
$$\int\limits_{x_l}^x dy e^{2i\varphi(y)+2i\theta(y)}v(y)\int_{y}^xds
e^{2i\xi(s)} v(s)=
{1\over 2} e^{4i\varphi(x_l)}\left(\int_{x_l}^xdy
e^{2i\xi(y)} v(y)\right)^2+ O(l^{1-\g}).$$
We use $\gamma$ as a general notations for universal positive constants,
they may change from line to line.
Combining (2.6), (2.8) one obtains
$$\ln{ R(x_{l+1})\over R(x_l)}={1\over 2\pi l}\im
\left(e^{2i\varphi(x_l)}\int\limits_{x_l}^{x_{l+1}} dy  
e^{2i\xi(y)} v(y)\right)
+{1\over 8\pi^2 l^2}\left|\int_{x_l}^{x_{l+1}}dy
e^{2i\xi(y)} v(y)\right|^2$$
$$+{1\over 8\pi^2 l^2}\re\left[
e^{4i\varphi(x_l)}\left(\int_{x_l}^{x_{l+1}}dy
e^{2i\xi(y)} v(y)\right)^2\right]+O(l^{-1-\g}).\tag 2.9$$
In a similar way,
$$\varphi(x_{l+1})-\varphi(x_l)=
{1\over 2}\re\left(\int\limits_{x_l}^{x_{l+1}} 
dy  e^{2i\theta} vp^{-1}\right)
+O(l^{-2})$$
$$=
{1\over 2\pi l}\re 
\left (e^{2i\varphi(x_l)}\int\limits_{x_l}^{x_{l+1}} dy  
e^{2i\xi} v\right) $$
$$-{1\over 8\pi^2 l^2}\im\left[
e^{4i\varphi(x_l)}\left(\int_{x_l}^xdy
e^{2i\xi(y)} v(y)\right)^2\right]-{1\over 4\pi^2 l^2}\im I_l(E)
+ O(l^{-1-\g}),\tag 2.10$$
where 
$$I_l(E)=\int\limits_{x_l}^{x_{l+1}}dy\int\limits^{x_{l+1}}_y ds 
e^{2i(\xi(s)-\xi(y))}
v(s)v(y).\tag 2.11$$
The basic properties of $I_l(E)$ are described by the following lemma.
\proclaim{Lemma 2.2}
For $v\in L_1({\Bbb T})$, $I_l(E)$ admits a representation of the form 
$$ I_l(E)={\Cal I}_l(E)+ O(l^{1-\g}),\quad l\rightarrow \infty,$$
where ${\Cal I}_l(E)$ is a $C^1$ function of $E$,
satisfying the estimates
$${\Cal I}_l(E)=
O(l),\quad  {d\over dE}{\Cal I}_l(E)= O(l).$$
\endproclaim
The proof of this lemma is given in appendix 1.

\par To derive a closed system for $R(x_l)$, $\theta(x_l)$
we need the following refinement of  lemma 2.1.
\proclaim{Lemma 2.3}
Let $v\in L_1({\Bbb T})$ and satisfy
$$|\hat v^{(k)}(n)|\leq C<n>^{-k},\quad k=1,2,3,\tag 2.12$$
where
$$\hat v^{(k)}(n)=\hat v^{(k-1)}(n)-\hat v^{(k-1)}(n-1),
\quad \hat v^{(0)}=\hat v.$$
Then,
$$\int\limits_{x_l}^{x_{l+1}} dye^{2i\xi(y)}v(y)
=e^{2i\omega_l}\pi \left({2l\over F}\right)^{1/2}\hat v(l)
+t_{l+1}-t_l+O(l^{-\gamma}),$$
where
$$\omega_l=
-{\pi^3l^3\over 3F}+\pi l(\k^\prime\ln l+E^\prime)+{\pi\over 8},$$
$$\k^\prime={2\k\over F},\quad
E^\prime={E-2\k+\k\ln\left(\pi^2\over F\right)\over F}.$$
$\{t_l\}$ is a bounded sequence: $|t_l|\leq C$.
\endproclaim
See appendix 1 for the proof.
\par Representations (2.9), (2.10) together with above lemma
give:
$$ 
\ln{ R(x_{l+1})\over R(x_l)}={1\over \sqrt{2lF}}
\im\left(e^{2i\omega_l+2i\varphi(x_l)}
\hat v(l)\right)+
{1\over 4lF}
\re\left(e^{4i\omega_l+4i\varphi(x_l)}
\hat v^2(l)\right)$$
$$+{1\over 4lF}|\hat v(l)|^2+
{1\over 2\pi}\im\left(
e^{2i\varphi(x_{l+1})}{t_{l+1}\over l+1}-
e^{2i\varphi(x_{l})}{t_{l}\over l}\right)+O(l^{-1-\g}),\tag 2.13$$
$$
\varphi(x_{l+1})-\varphi(x_{l})=
{1\over \sqrt{2lF}}
\re\left(e^{2i\omega_l+2i\varphi(x_l)}
\hat v(l)\right)-
{1\over 4lF}
\im\left(e^{4i\omega_l+4i\varphi(x_l)}
\hat v^2(l)\right)$$
$$-{1\over 4\pi^2 l^2}\im {\Cal I}_l(E)+
{1\over 2\pi}\re\left(
e^{2i\varphi(x_{l+1})}{t_{l+1}\over l+1}-
e^{2i\varphi(x_{l})}{t_{l}\over l}\right)+O(l^{-1-\g}).
\tag 2.14$$

\par Assume now that ${F\over \pi^2}\in{\Bbb Q}$:
$${\pi^2\over F}={3p\over q},\quad p,q\in {\Bbb N}.$$
We will consider $R(x_l)$, $\varphi(x_l)$ along the subsequence
$l=qk$. Set
$$\tilde R(k)=R(x_{kq}),\quad \tilde\varphi(k)= \varphi(x_{kq}).$$
Then $\tilde R$, $\tilde\varphi$ solve the system:
$$
\ln{ \tilde R(k+1)\over \tilde R(k)}=
\im\left(e^{2i\Omega(k)+2i\tilde\varphi(k)}
b(k)\right)+
{1\over 2}\re\left(e^{4i\Omega(k)+4i\tilde\varphi(k)}
b^2(k)\right)$$
$$+{1\over 2}|b(k)|^2
+
{1\over 2\pi q}\im\left(
e^{2i\tilde\varphi(k+1)}{t_{q(k+1)}\over k+1}-
e^{2i\tilde\varphi(k)}{t_{qk}\over k}\right)+O(k^{-1-\g}),\tag 2.15$$
$$\tilde\varphi(k+1)-\tilde\varphi(k)=
\re\left(e^{2i\Omega(k)+2i\tilde\varphi(k)}
b(k)\right)-
{1\over 2}
\im\left(e^{4i\Omega(k)+4i\tilde\varphi(k)}
b^2(k)\right)$$
$$-b_1(k)-
\im \tilde {\Cal I}_k(E)$$
$$+
{1\over 2\pi}\re\left(
e^{2i\tilde\varphi(k+1)}{t_{q(k+1)}\over k+1}
-
e^{2i\tilde\varphi(k)}{t_{qk}\over k}\right)
+O(k^{-1-\g}).
\tag 2.16$$
Here
$$\Omega(k)=\pi kq(E^\prime +\k^\prime\ln(kq)),$$
$$b(k)={1\over \sqrt{2Fqk}}\hat v(qk)w(s(k)),\quad
b_1(k)={1\over 2Fqk}|\hat v(kq)|^2\im w_1(s(k)),$$
$$
s(k)={d\over dk}\Omega(k)=\pi q(E^\prime +\k^\prime+\k^\prime\ln(kq)),$$
$$
w(s)=e^{i\pi/4}\sum\limits_{r=0}^{q-1}e^{-2\pi i{p\over q}r^3+
2 i{s\over q}r},$$
$$w_1(s)=
\sum\limits_{r=1}^{q-1}e^{-2\pi i{p\over q}r^3+
2i{s\over q}r}\sum\limits_{r_1=0}^{r-1}e^{2\pi i{p\over q}r_1^3-
2 i{s\over q}r_1},\quad \roman{if}\,\,q>1,$$
 and $w_1(k)\equiv 0$ if $q=1$,
$$
 \tilde {\Cal I}_k(E)={1\over 4\pi^2 q^2k^2}
\sum\limits_{l=kq}^{(k+1)q-1}{\Cal I}_l(E).$$

\subsubhead
 2.1.3. Case of 
$\k=0$
\endsubsubhead
\par In this case (2.15) gives for $K_2>K_1$ sufficiently large
$$\ln{ \tilde R(K_2)\over \tilde R(K_1)}=
\im\left(\sum_{k=K_1}^{K_2-1}e^{2i\pi E^\prime qk+2i\tilde\varphi(k)}
b(k)\right)+{1\over 2}\re\left(\sum_{k=K_1}^{K_2-1}
e^{4i\pi E^\prime qk+4i\tilde\varphi(k)}
b^2(k)\right)
$$
$$
+{1\over 2}\sum_{k=K_1}^{K_2-1}|b(k)|^2+O(K_1^{-\g}).\tag 2.17$$
Consider the second sum
$$\sum_{k=K_1}^{K_2-1}
e^{4i\pi E^\prime qk+4i\tilde\varphi(k)}
b^2(k)\tag 2.18$$
Summing by parts, one gets
$$(2.18)={1\over e^{-4\pi i E^\prime q}-1}
\sum_{k=K_1}^{K_2-1}e^{4i\pi E^\prime qk}\left[e^{4i\tilde\varphi(k+1)}
b^2(k+1)-e^{4i\tilde\varphi(k)}b^2(k)\right]$$
$$+O(K_1^{-1/2})
=O(K_1^{-1/2}),\tag 2.19$$
provided $2E^\prime q\not\in {\Bbb Z}$.
\par In a similar way, one has
$$\sum_{k=K_1}^{K_2-1}e^{2i\pi E^\prime qk+2i\tilde\varphi(k)}
b(k)$$
$$=2i{1\over e^{-2\pi i E^\prime q}-1}
\sum_{k=K_1}^{K_2-1}e^{2i\pi E^\prime qk+2i\tilde\varphi(k)}b(k)
\re\left(e^{2i\pi E^\prime qk+2i\tilde\varphi(k)}
b(k)\right)+O(K_1^{-1/2})$$
$$
=i{1\over e^{-2\pi i E^\prime q}-1}
\sum_{k=K_1}^{K_2-1}|b(k)|^2+O(K_1^{-1/2}).$$
In particular,
$$\im \left (\sum_{k=K_1}^{K_2-1}e^{2i\pi E^\prime qk+2i\tilde\varphi(k)}
b(k)\right)=
-{1\over 2}\sum_{k=K_1}^{K_2-1}|b(k)|^2+O(K_1^{-1/2}).\tag2.20$$
Combining (2.17), (2.19), (2.20),
$$\ln{ \tilde R(K_2)\over \tilde R(K_1)}=O(K_1^{-\g}).$$
This means that there exists
$$\lim\limits_{k\rightarrow +\infty}\tilde R(k)=R_\infty,\quad
0<R_\infty<\infty,$$
 which together with (2.4), (2.7) allows to conclude
that any solution $\psi$ of (2.1) satisfies for sufficiently large N
$$ C_1(\psi)N^{1/2}\leq \int_0^Ndx|\psi|^2 \leq C_2(\psi)N^{1/2},$$
with some constants $C_1(\psi),\,C_2(\psi)$ depending on $\psi$.
Therefore,  for $2E^\prime q\not\in {\Bbb Z}$ all solutions have
the same rate of $L_2$ norm growth as $N\rightarrow +\infty$,
and there is no subordinate solution on the right.
By Gilbert-Pearson theory [12, 13],
this implies that 
the singular continous spectrum of $H$
is empty, the  point spectrum is contained in the set
$\{E:2E^\prime q\in {\Bbb Z}\}$ and
$$\Sigma_{ac}(H)={\Bbb R}.$$
Notice also that for $\k=0$, $H$ is unitary equivalent to
$H+F$.

\subhead
2.2. Case $\k\neq 0$: reduction to a model system
\endsubhead
\subsubhead
2.2.1. Adiabatic regime
\endsubsubhead

\par Since $s(k)$, $b(k)$, $b_1(k)$ are slowly
varying functions of $k$ equations  (2.15), (2.16)
can be treated adiabatically except for  relatively small
vicinities of the stationary points $K_m$
defined by the equation
$$\Omega^\prime(K_m)\equiv 
s(K_m)=\pi m,\quad m\in {\Bbb Z},$$
which  means
$$K_m=A\Lambda^m,\quad \Lambda=e^{1\over q\k^\prime},\quad 
A={1\over q}e^{-{E^\prime +\k^\prime\over \k^\prime}}.\tag 2.21$$
For the sake of definiteness we will assume that
$\k>0$. So, it is the limit $m\rightarrow +\infty$ that we 
will be interested in.
\par We are going to study $\tilde R(k)$, $\tilde \varphi(k)$
along the subsequence $k_m$,
$k_m=[\tilde k_m]$, where $\tilde k_m$ solves
the equation 
$$s(\tilde k_m)=\pi (m-{1\over 2}).$$
Define
$$ \hat K_m=[K_m],\quad K_m^\pm=[K_m\pm K_m^{1-\eta}],$$
where $0<\eta<{1\over 2}$ to be  fixed later.
\par First we consider the intervals $J_m^\pm$,
$J_m^-=[k_m, K_m^-]$, $J_m^+=[K^+_m, k_{m+1}]$.
Clearly,  
for $k\in J_m^-\cup  J_m^+$,
$$\Omega^{(1)}(k)\equiv \Omega(k)-\Omega(k-1)=s(k)+O(k^{-1})$$
 satisfies
$$|1-e^{-2i\Omega^{(1)}(k)}|\geq CK_m^{-\eta},$$
provided
$m$ is  sufficiently large. 
\par 

Let us rewrite (2.15), (2.16) in the form
$$\ln{ \tilde R(K_2+1)\over \tilde R(K_1)}=
\im\left(\sum_{k=K_1}^{K_2}
e^{2i\Omega(k)+2i\tilde\varphi(k)}
b(k)\right)+
{1\over 2}\re\left(\sum_{k=K_1}^{K_2}e^{4i\Omega(k)+4i\tilde\varphi(k)}
b^2(k)\right)$$
$$+\sum_{k=K_1}^{K_2}{1\over 2}|b(k)|^2
+O(K_m^{-\g}),\tag 2.22$$
$$\tilde\varphi(K_2+1)-\tilde\varphi(K_1)=
\re\left(\sum_{k=K_1}^{K_2}e^{2i\Omega(k)+2i\tilde\varphi(k)}
b(k)\right)-
{1\over 2}
\im\left(\sum_{k=K_1}^{K_2}e^{4i\Omega(k)+4i\tilde\varphi(k)}
b^2(k)\right)$$
$$-\sum_{k=K_1}^{K_2}b_1(k)-
\sum_{k=K_1}^{K_2-1}\im \tilde {\Cal I}_k(E)+O(K_m^{-\g}),\tag 2.23$$
$[K_1,K_2]\subset J_m^-\cup  J_m^+$.
Consider the sum
$$\sum\limits_{k=K_1}^{K_2}e^{2i\Omega(k)+2i\tilde\varphi(k)}
b(k)\tag 2.24$$
Summing by parts one gets
$$(2.24)=
\sum\limits_{k=K_1}^{K_2}e^{2i\Omega(k)}
\bigg[(e^{-2i\Omega^{(1)}(k+1)}-1)^{-1}e^{2i\tilde\varphi(k+1)}b(k+1)$$
$$-
(e^{-2i\Omega^{(1)}(k)}-1)^{-1}e^{2i\tilde\varphi(k)}b(k)\bigg]
+
O(K_m^{-1/2+\eta})$$
$$=i\sum\limits_{k=K_1}^{K_2}
|b(k)|^2(e^{-2is(k)}-1)^{-1}
+
i\sum\limits_{k=K_1}^{K_2}e^{4i\Omega(k)+4i\tilde\varphi(k)}
b^2(k)(e^{-2is(k)}-1)^{-1}$$
$$+\sum\limits_{k=K_1}^{K_2}e^{2i\Omega(k)+2i\tilde\varphi(k)}
b(k)\left((e^{-2is(k+1)}-1)^{-1}-
(e^{-2is(k)}-1)^{-1}\right)+O(K_m^{-1/2+\eta}).\tag 2.25$$
The last sum in (2.25) can be estimated as folows.
$$\left |\sum\limits_{k=K_1}^{K_2}e^{2i\Omega(k)+2i\tilde\varphi(k)}
b(k)\left((e^{-2is(k+1)}-1)^{-1}-
(e^{-2is(k)}-1)^{-1}\right)\right|$$
$$\leq C\sum\limits_{k=K_1}^{K_2}k^{-3/2}|e^{-2is(k)}-1|^{-2}
\leq C\left(\int_{K_1}^{K_2}dy y^{-3/2}|e^{-2is(y)}-1|^{-2}+
K_m^{-3/2+3\eta}\right)$$
$$\leq CK_m^{-1/2+\eta}.\tag 2.26$$
Next we consider the sums 
$$\sum\limits_{k=K_1}^{K_2}e^{4i\Omega(k)+4i\tilde\varphi(k)}f(k),
\tag 2.27$$
where $k_m\leq K_1\leq K_2\leq k_{m+1}$, and $f(k)$ is either
$b^2(k)(e^{-2i\Omega^{(1)}(k-1)}-1)^{-1}$
or $b^2(k)$.
If $K_1$, $K_2$ satisfy
$k_m+K_m^{1-\eta}\leq K_1\leq K_2\leq K_m^-$ or
$K_m^+\leq K_1\leq K_2\leq k_{m+1}-K_m^{1-\eta}$
then
$$|1-e^{-4i\Omega^{(1)}(k)}|\geq CK_m^{-\eta}.$$

Proceeding in the same way as in (2.25)
one gets
$$|(2.27)|\leq C\left(\sum\limits_{k=K_1}^{K_2}\left(
k^{-3/2}||e^{-2is(k)}-1|^{-1}|e^{-4is(k)}-1|^{-1}\right.\right.$$
$$+\left.\left.
k^{-2}
|e^{-2is(k)}-1|^{-3}+k^{-2}|e^{-4is(k)}-1|^{-3}\right)+
K_m^{-1+2\eta}\right)$$
$$\leq C\left(\int_{K_1}^{K_2} dy y^{-2}\left(|e^{-2is(y)}-1|^{-3}+
|e^{-4is(y)}-1|^{-3}\right)+K_m^{-1/2+\eta}\right)$$
$$\leq CK_m^{-1/2+\eta}.$$
 On the other hand if $k_m\leq K_1\leq K_2\leq k_m+K_m^{1-\eta}$,
or  $k_{m+1}-K_m^{1-\eta}\leq K_1\leq K_2\leq k_{m+1}$, then
$$\sum\limits_{k=K_1}^{K_2}e^{4i\Omega(k)+4i\tilde\varphi(k)}
f(k)=O(K_m^{-\eta}). $$
Therefore, one has
$$\sum\limits_{k=K_1}^{K_2}e^{4i\Omega(k)+4i\tilde\varphi(k)}
f(k)=O(K_m^{-\g}),\tag 2.28$$
for any $K_1$, $K_2$ such that
$[K_1,\, K_2]\subset {\Cal I}_m^-\cup  {\Cal I}_m^+$.
\par Combining (2.23), (2.24), (2.26)  we get
$$\sum\limits_{k=K_1}^{K_2}e^{2i\Omega(k)+2i\tilde\varphi(k)}
b(k)=i \sum\limits_{k=K_1}^{K_2}
|b(k)|^2(e^{-2is(k)}-1)^{-1}+O(K_m^{-\g}).\tag 2.29$$
Representations (2.15), (2.16), (2.26), (2.27) allow to conclude that
for any $K_1$, $K_2$, $[K_1,\, K_2]\subset J_m^-\cup  J_m^+$
$$\ln{ \tilde R(K_2+1)\over \tilde R(K_1)}=O(K_m^{-\g}),\tag 2.30$$
$$\tilde \varphi(K_2+1)-\tilde \varphi(K_1)=
-{1\over 2}\sum\limits_{k=K_1}^{K_2}|b(k)|^2\cot s(k)
-\sum\limits_{k=K_1}^{K_2}(b_1(k)+\im\tilde I_k(E))
+O(K_m^{-\g}).\tag 2.31$$
The sums in the r.h.s. of (2.31) allow some further simplifications. 
One has
$$\sum\limits_{k=K_1}^{K_2}|b(k)|^2\cot s(k)$$
$$=
{|v_0|^2\over 2Fq}
\int_{K_1}^{K_2}
dy y^{-1}|w(s(y)|^2(\ln (qy) )^{-2\beta}\cot s(y)+O(K_m^{-\g})$$
$$=|b_0|^2\int_{s(K_1)}^{s(K_2)}
ds|w(s)|^2(s-s_0)^{-2\beta}\cot s+O(K_m^{-\g}),\tag 2.32$$
where 
$$b_0={v_0\over \sqrt{2 Fq}}\mu ^{-1/2+\beta},\quad
\mu=\pi q\kappa^\prime,
\quad s_0=\pi q(E^\prime+\kappa^\prime).$$
In a similar way,
$$\sum\limits_{k=K_1}^{K_2}b_1(k)=
|b_0|^2\int_{s(K_1)}^{s(K_2)}
ds\im w_1(s)(s-s_0)^{-2\beta}+O(K_m^{-\g}).\tag2.33$$
(2.31), (2.32), (2.33) give
$$\tilde \varphi(K_m^-)-\tilde \varphi(k_m)=
-{|b_0|^2\over 2}|w(\pi m)|^2(\pi m-s_0)^{-2\beta}
\ln(\mu K_m^{-\eta})$$
$$-{|b_0|^2\over 2}\int_{\pi(m -1/2)}^{\pi m}
ds\cot s\big(|w(s)|^2(s-s_0)^{-2\beta}
-|w(\pi m)|^2(\pi m-s_0)^{-2\beta}\big)$$
$$-|b_0|^2\int_{\pi(m-1/2)}^{\pi m}
ds\im w_1(s)(s-s_0)^{-2\beta}-
\sum\limits_{k=k_m}^{K_m^-}\im \tilde I_k(E)+O(K_m^{-\g}),
\tag 2.34$$
$$\tilde \varphi(k_{m+1})-\tilde \varphi(K_m^+)=
{|b_0|^2\over 2}|w(\pi m)|^2(\pi m-s_0)^{-2\beta}
\ln(\mu K_m^{-\eta})$$
$$
-{|b_0|^2\over 2}\int_{\pi m+\mu K_m^{-\eta}}^{\pi (m +1/2)}
ds\cot s
\big(|w(s)|^2(s-s_0)^{-2\beta}-
|w(\pi m)|^2(\pi m-s_0)^{-2\beta}\big)$$
$$-|b_0|^2\int^{\pi(m+1/2)}_{\pi m}
ds\im w_1(s)(s-s_0)^{-2\beta}-
\sum\limits^{k=k_{m+1}}_{k=K_m^+}\im \tilde I_k(E)+O(K_m^{-\g}).
\tag 2.35$$

\subsubhead
2.2.2. Vicinities of stationary points
\endsubsubhead
In this subsection we analyse  
 system (2.15), (2.16)
in the $K_m^{1-\eta}$- vicinity of the turning point $K_m$:
$$k\in \delta_m,\quad \delta_m=[K_m^-,K_m^+].$$
In this vicinity one has
$$b(k)e^{2i\Omega(k)}=\left({\mu\over K_m}\right)^{1/2}d_m
e^{2i\Phi_0(m)+i\mu{(k-K_m)^2\over K_m}}
+O(K_m^{1-3\eta}),$$
$$d_m=b_0w(\pi m)
(\pi m-s_0)^{-\beta},\quad\Phi_0(m)=\Omega(K_m)-\pi m K_m,
\tag 2.36$$
provided ${1\over 3}<\eta<{1\over 2}$. 
By (2.21),
$$\Phi_0(m)=\rho\Lambda^m,$$
where
$$
\rho=-
\pi\kappa^\prime e^{-{E^\prime+\kappa^\prime\over \kappa^\prime}},$$
one can  consider $\rho$ as a new spectral parameter. 
\par From now on we fix $\eta$ in such a way that
$${3\over 8}<\eta<{1\over 2}.\tag 2.37$$
\par Introduce the vectors $ \chi (k)\in {\Bbb C}^2$,
$$ \chi (k)=\tilde R(k)e^{i\tilde \varphi(k)\s_3}
{1\choose 1},\quad \s_3=\pmatrix 1&0\cr 0&-1\cr\endpmatrix.$$
Notice that 
$$\chi(k)=e^{-i\xi(x_{qk})\s_3}\left(Q_\xi(\xi(x_{qk})){1\choose 1}+
Q(\xi(x_{qk})){1\choose -1}\right).$$
For $k\in \delta_m$, (2.15), (2.16) imply
$$ \chi (k+1)= \chi (k) +i
\pmatrix  \bar b(k)e^{-2i\Omega(k)-2i\tilde\varphi(k)}&0\cr
0&b(k)e^{2i\Omega(k)+2i\tilde\varphi(k)}\cr\endpmatrix \chi (k) +
O(k^{-1}) \chi (k)$$
$$=(A_0(k,m)+A_1(k,m,\chi)) \chi (k),\tag 2.38$$
where
$$A_0=
I+\left({\mu\over K_m}\right)^{1/2}
D_me^{i\mu{(k-K_m)^2\over K_m}\s_3},\quad
D_m=i\pmatrix 0&\bar d_me^{-2i\Phi_0(m)}
\cr- d_me^{2i\Phi_0(m)}&0\cr\endpmatrix,$$
and 
$$A_1=O(K_m^{1-3\eta}).\tag 2.39$$
\par On the interval $\delta_m$
system (2.38) can be approximated by the differential equation
$${d\over dk}\chi=\left({\mu \over K_m}\right)^{1/2}D_m
e^{i\mu{(k-K_m)^2\over K_m}\s_3}\chi.\tag 2.40$$
Set
$$\chi(k)=\psi(y),\quad y=\left({\mu \over K_m}\right)^{1/2}
(k-K_m).$$
Then (2.40) takes the form
$${d\over dy}\psi=D_me^{i y^2\s_3}\psi.\tag 2.41$$
Note that if $\psi(y)$ is a solution 
then $\s_1\overline{\psi(y)}$,
$\s_3\psi(y)$ also satisfy (2.41).
Here 
$\s_1=\pmatrix 0&1\cr 1&0\cr\endpmatrix$.
\par Since $d_m$ does not depend on $y$,
(2.41) can be solved explicitely in terms of 
Hermite functions. One checks directly that
$$\psi(y)={H_{\lambda_m}(e^{-i\pi/4}y)\choose
-d_m
e^{2i\Phi_0(m)+i y^2-i\pi/4}H_{\lambda_m-1}(e^{-i\pi/4}y)},
\quad \lambda_m=-{i\over 2}|d_m|^2,\tag 2.42$$
is a solution of (2.41).
Here $H_\lambda(z)$ stands for the standard
Hermite function: it satisfies the equation
$$f_{zz}-2z f_z+2\lambda f=0,$$
and
$$
H_\lambda(z)=(2z)^\lambda(1+O(z^{-2})),\quad z\rightarrow \infty,$$
$-\pi/2<\arg z\leq \pi/2$.
\par Let us introduce the matrix solutions $\Psi^\pm(y)$:
$$\Psi^+(y)=(\psi(y),\s_1\overline{\psi(y)}),\quad
\Psi^-(y)=\s_3\Psi^+(-y)\s_3.\tag 2.43$$
As $y\rightarrow +\infty$,
$$\Psi^+(y)=
e^{-{\pi |d_m|^2\over 8}}(2y)^{\l_m\s_3}+O(y^{-1}),
\tag 2.44$$
uniformely with respect to $m$ sufficiently large.
\par The determinat of $\Psi^\pm(y)$
does not depend on $y$:
$$\det \Psi^\pm=e^{-{\pi |d_m|^2\over 4}}.$$ 
\par Using $\Psi^-$ one can rewrite 
 full equation (2.38) in the form
$$\chi(k)=\Psi^-(y)a_-(k),\quad y=\left({\mu \over K_m}\right)^{1/2}
(k-K_m)\tag 2.45$$
$$a_-(k)=a(K_m^-)+\sum\limits_{j=K_m^-}^{k-1}A_2(j,m)a_-(j),
\quad k\geq K_m^-+1,
\tag 2.46$$
$$A_2(k,m)=
\left(\Psi^-\left(y+\left({\mu\over K_m}\right)^{1/2}\right)\right)^{-1}
\bigg[\Psi^-(y)
+
\left({\mu\over K_m}\right)^{1/2}{d\over dy}\Psi^-(y)
$$
$$-
\Psi^-\left(y+\left({\mu\over K_m}\right)^{1/2}\right)
+A_1(k,m)\Psi^-(y)\bigg].$$
It folows directly from  (2.40), (2.41) that
$$|\Psi^\pm|\leq C,\quad \left|{d\Psi^\pm\over dy}\right| \leq C, \quad 
\left|{d^2\Psi^\pm\over dy^2}\right|\leq C<y>, \quad y\in {\Bbb R},
\tag 2.47$$
which together with (2.39) implies
$$|A_2|
\leq CK_m^{1-3\eta},\quad k\in \delta_m.\tag 2.48$$
(2.46), (2.48) allow to conclude
$$|a_-(k)-a_-(K_m^-)|
\leq CK_m^{-\g}\max\limits_{k\in \delta_m}|a_-(k)| ,
\quad k\in \delta_m,\tag 2.49$$
provided (2.37) is satisfied.
This means that for $m$ sufficiently large
$$|a_-(k)-a_-(K_m^-)|\leq CK_m^{-\g}a_-(K_m^-),
\quad k\in \delta_m,\tag 2.50$$
\par In a similar way, setting
$$\chi(k)=\Psi^+(y)a_+(k),\tag 2.51$$
one gets
$$|a_+(k)-a_-(K_m^+)|\leq CK_m^{-\g}|a_-(K_m^+)|,
\quad k\in \delta_m.\tag 2.52$$
Comparing (2.46), (2.51) 
and taking into account (2.44), (2.47), (2.50), (2.52)
we obtain
$$\chi(K_m^+)=(2\sqrt{\mu}K_m^{1/2-\eta})^{\l_m\s_3}\Psi_+^{-1}(0)
\s_3\Psi_+(0)\s_3(2\sqrt{\mu}K_m^{1/2-\eta})^{-\l_m\s_3}\chi(K_m^-)$$
$$+O(K_m^{-\g})\tilde R(K_m^-).\tag 2.52$$
By (2.42) the expression $\Psi_+^{-1}(0)
\s_3\Psi_+(0)\s_3$ may be represented as
$$\Psi_+^{-1}(0)
\s_3\Psi_+(0)\s_3=e^{-i\Phi_0(m)\s_3}S(d_m)e^{i\Phi_0(m)\s_3},$$
$$S(\xi)=S^{-1}_0(\xi)\s_3S_0(\xi)\s_3,
$$
$$
S_0(\xi)=\pmatrix H_\l(0)&-e^{i\pi/4}\bar\xi H_{-\l-1}(0)\cr
-e^{-i\pi/4}\xi H_{\l-1}(0)& H_{-\l}(0)\cr\endpmatrix,
\quad \l=-{i\over 2}|\xi|^2.\tag 2.53$$
Clearly,
$$S(\xi)=\pmatrix s(\xi)& \bar r(\xi)\cr
                  r(\xi)&s(\xi)\cr\endpmatrix,
\quad s^2(\xi)-|r(\xi)|^2=1,$$
$$r(\xi)=
2e^{-i\pi/4+\pi|\xi|^2/4} H_\l(0) H_{\l-1}(0),
\quad s(\xi)= e^{\pi|\xi|^2/4} (|H_\l(0)|^2+|\xi|^2|H_{\l-1}(0)|^2).$$
As a consequence,
$$r(d_m)=r_0w(\pi m)m^{-\b}(1+O(m^{-2\b})),\quad s(d_m)=1+O(m^{-2\b}),\quad
r_0=e^{-i\pi/4}\sqrt{\pi}b_0,$$
$${\partial\over \partial \rho}r(d_m)=O(m^{-1-2\b}).$$

\par (2.52)
together with (2.30), (2.34), (2.35) leads to a ``closed
system'' for $\hat \chi(m)=\chi(k_m)$:
$$\hat \chi(m+1)={\Cal A}(m)\hat \chi(m)+
O(e^{-m\g}|\hat \chi(m)|),\tag 2.54$$
$${\Cal A}(m)=
e^{-i(\Phi_0(m)+\triangle_+(m))\s_3}
S(d_m)e^{-i(\Phi_0(m)+\triangle_-(m))\s_3},$$
where
$$\triangle_\pm(m)=\Phi_1(m,\rho)+\Phi_2^\pm(m,\rho)+
\Phi_3^\pm(m,\rho),$$
$$\Phi_1(m,\rho)=-{|d_m|^2\over 4}
\ln\left(\mu\over 4K_m\right),$$
$$\Phi_2^-(m,\rho)=-{|b_0|^2\over 2}\int_{\pi(m -1/2)}^{\pi m}
ds\cot s\big(|w(s)|^2(s-s_0)^{-2\beta}
-|w(\pi m)|^2(\pi m-s_0)^{-2\beta}\big)$$
$$-|b_0|^2\int_{\pi(m-1/2)}^{\pi m}
ds\im w_1(s)(s-s_0)^{-2\beta},$$
$$\Phi_2^+(m,\rho)={|b_0|^2\over 2}
\int_{\pi m}^{\pi (m +1/2)}
ds\cot s
\big(|w(s)|^2(s-s_0)^{-2\beta}-
|w(\pi m)|^2(\pi m-s_0)^{-2\beta}\big)$$
$$+|b_0|^2\int^{\pi(m+1/2)}_{\pi m}
ds\im w_1(s)(s-s_0)^{-2\beta}$$
$$\Phi_3^-(m,\rho)=-\sum\limits_{k=k_m}^{\hat K_m}\im\tilde I_k(E),$$
$$\Phi_3^+(m,\rho)=
\sum\limits^{k=k_{m+1}}_{k=\hat K_m}\im\tilde I_k(E).$$

Note that $\Phi_1,\Phi_2^\pm$ are smooth functions of $\rho$ 
satisfying
$$
|\Phi_1(m,\rho)|\leq Cm^{1-2\b},\quad \left|{\partial\over \partial\rho}
\Phi_1(m,\rho)\right|\leq Cm^{-2\b},\tag 2.55$$
$$|\Phi_2^\pm(m,\rho)|\leq Cm^{-2\b},\quad \left|{\partial\over \partial\rho}
\Phi_2^\pm(m,\rho)\right|\leq Cm^{-1-2\b}.\tag 2.56$$
It follows from lemma 2.2 that
$$|\Phi_3^\pm(m,\rho)|\leq C\tag 2.57$$
$$ \left|
\Phi_3^\pm(m,\rho)-\Phi_3^\pm(m;\rho^\prime)\right|\leq C(
\Lambda ^{-m}+|\rho-\rho^\prime|).\tag 2.58$$
Fix a function $\varphi\in C_0^\infty ({\Bbb R})$
such that $\varphi\geq 0$, $\int_{\Bbb R} dx\varphi(x)=1$.
Set
$$\tilde\Phi_3^\pm(m,\rho)=\kappa_m\int_{\Bbb R}dy
\varphi(\kappa_m(y-\rho))\Phi_3^\pm(m,y),$$
where $\kappa_m=\Lambda_1^{-m},$
$1<\Lambda_1<\Lambda$ to be fixed later.
Then (2.58) implies
$$|\Phi_3^\pm(m,\rho)-\tilde\Phi_3^\pm(m,\rho)|\leq Ce^{-\g m},$$
$$\left|{\partial\over \partial\rho}
\tilde\Phi_3^\pm(m,\rho)\right|\leq C\Lambda_1^{m}.\tag 2.59$$
Therefore, one can rewrite (2.54) in the form
$$\hat \chi(m+1)={\Cal A}_0(m)\hat \chi(m)+
O(e^{-m\g}|\hat \chi(m)|),\tag 2.60$$
$${\Cal A}_0(m)=e^{-i\Gamma_+(m)\s_3}S(d_m)e^{i\Gamma_-(m)\s_3},$$
where
$$\Gamma_\pm(m)=\rho \Lambda^m+\Gamma_1^\pm(m),\quad
\Gamma_1^\pm(m)=
\Phi_1(m,\rho)+
\Phi^\pm_2(m,\rho)+\tilde\Phi^\pm_3(m,\rho).$$
As a simple consequence of (2.60)(=(2.54)) one obtains
\proclaim{Proposition 2.1}
The operator H has no point spectrum.
\endproclaim
\par {\it Proof.}  (2.60) implies
$$\tilde R(k_{m+1})\geq (1-Cm^{-\b})\tilde R(k_{m}).$$
As a consequence, one gets
$$ \tilde R(k_m)\geq C(\psi)
e^{-Cm^{1-\beta}}.$$
Combining this estimate with (2.51), (2.52), (2.30), (2.13), (2.7), (2.4)
one can check easily that for any solution $\psi$

$$\int_0^N dx |\psi|^2\geq C(\psi)N^{1/2}e^{-C(\ln N)^{1-\b}}.$$
Therefore, $\psi$  can not belong to $L_2$.\hfill $\blacksquare$

\head
3. Analysis  of the model system
\endhead
In this section  we study the model system
$$ \chi(m+1)={\Cal A}_0(m) \chi(m).\tag 3.1$$
\subhead
3.1. Positivity of the Lyapounov exponent
\endsubhead
\subsubhead
3.1.1. Pr\"ufer coordinates for {\rm (3.1)}
\endsubsubhead
We denote by $\chi_\a(m)$ the solution of (3.1)
satisfying
$$\chi_\a(M_0)=e^{i\a\s_3}{1\choose 1},\quad \a\in [0,\pi),$$
$M_0$ is supposed to be a large fixed number.
\par Define the Pr\"ufer variables $R_\a(m)$, $\varphi_\a(m)$ by
$$\chi_\a(m)=R_\a(m)e^{i\varphi_\a(m)\s_3}{1\choose 1}.$$
They solve
$$ 
{R_\a(m+1)\over R_\a(m)}=
|s(d_m)+r(d_m)e^{2i\Gamma_-(m)}\zeta(m)|,\tag 3.2$$
$$
{\zeta(m+1)\over \zeta(m)}=e^{2i\triangle \Gamma(m)}
{s(d_m) +\bar r(d_m)\bar \zeta(m)e^{-2i\Gamma_-(m)}\over 
s(d_m) +r(d_m)\zeta(m)e^{2i\Gamma_-(m)}},
\tag 3.3$$
where
$$\triangle \Gamma(m)=\Gamma_-(m)-\Gamma_+(m),\quad
\zeta(m)=e^{2i\varphi_\a(m)}.$$
From (3.2) we have
$$\ln R_\a(M)
={1\over 2} |r_0|^2 n(M)(1+O(n^{-\g}(M)))$$
$$+\re\left(
r_0\sum\limits_{m=M_0}^M m^{-\b}w(\pi m)e^{2i\Gamma_-(m)}\zeta(m)\right)
\tag3.4$$
$$-{1\over 2}
\re\left(r_0^2
\sum\limits_{m=M_0}^M m^{-2\b}w^2(\pi m)
e^{4i\Gamma_-(m)}\zeta^2(m)\right),\tag 3.5$$
where
$$n(M)=\sum\limits_{m=1}^Mm^{-2\b}|w(\pi m)|^2.$$
Note that 
$$w(\pi m)=w(\pi(m+q)),\quad \sum\limits_{j=0}^{q-1}|w(\pi j)|^2=q^2.$$
As a consequence,
$$n(M)=q\sum\limits_{m=M_0}^M m^{-2\beta}+ O(1)=
{q\over 1-2\beta}M^{1-2\b}+O(1),$$
as $M\rightarrow +\infty$.
\par Notice also that (3.3) implies
$$|\zeta_\rho^\prime(m+1)|\leq |\zeta_\rho^\prime(m)|
(1+Cm^{-\beta}) +Cm^{-\beta}\Lambda^m\leq Cm^{-\beta}\Lambda^m.\tag 3.6$$
Next two subsubsections are devoted to the proof of the following result.

\proclaim{Proposition 3.1}
For any $\a$  and for a.e. $\rho$ we have
$$\lim\limits_{m\rightarrow +\infty} {\ln R_\a(m)\over m^{1-2\b}}=r_*,
$$
where $r_*={|r_0|^2q\over 2(1-2\b)}$.
\endproclaim

\subsubhead
3.1.2. Estimates of sum  {\rm (3.5)}
\endsubsubhead

To prove  proposition 3.1 we are going to analyse
the sums
$$ \Sigma_1(M)=\sum\limits_{m=M_0}^M 
m^{-\b}w(\pi m)e^{2i\Gamma_-(m)}\zeta(m),\tag 3.7$$
$$  \Sigma_2(M) =  \sum\limits_{m=M_0}^M m^{-2\b}w^2(\pi m)
e^{4i\Gamma_-(m)}\zeta^2(m)\tag 3.8$$
and show that they are 
$o(n(M))$.
In this part of the paper
we follow closely the arguments
of [6, 8, 14]. We start by a technical lemma,
which is essentially lemma 10.1 of [6].
Consider the sums of the form
$$S_N(\rho)=
\sum\limits_{n=0}^N
a(n)f_{n}(L^{h+n}\rho)g(L^{h+n}\rho),$$
where $N\geq 1$,  $h\in{\Bbb R}$, $a(n)$ are real numbers,
$\{f_{n}\} \in C^1$, $\|f_{n}\|_\infty\leq 1$,  
$g(y)=\cos (ky+b)$.
We assume
$$\|f_{n}^\prime\|_\infty\leq K, $$
$$ |k|+|k|^{-1}\leq K,\quad L\geq \Lambda>1.$$

\proclaim{Lemma 3.1} 
One has
$$\sup\limits_{I, |I|=1}\int_I d\rho \exp(tS_N(\rho))
\leq 
e^{B(K,\Lambda)t^2A^2(N)+B(K,\Lambda)t{\Cal Q}(N)},
\tag 3.9$$
$$
A^2(N)=\sum\limits_{n=0}^N a^2(n),\quad
{\Cal Q}(N)=
\sum\limits_{n=0}^N|a(n)|(\|f_n^\prime\|_\infty+L^{-(h+n)}),$$
provided
$$0\leq t\max\limits_{
0\leq n\leq N}
|a(n)|\leq 1,$$
the supremum being taken over all intervals 
$I\subset {\Bbb R}$, $|I|=1$.
\endproclaim
Here and below $B(K,\Lambda)$ are positive constants that 
 depend only on
$K$, $\Lambda$, they may change from line
to line.

\par The proof of this lemma is given in appendix 2.

\par Applying (3.9)
to  sum  (3.8) one gets the following result.
Fix an interval $I\subset {\Bbb R}$, $|I|=1$.
\proclaim{Lemma 3.2}
 There exist $\varepsilon_0,\varepsilon_1>0$,
such that 
$$\mes 
\{ \rho \in I: |\Sigma_2(M)|\geq M^{1-2\beta-\varepsilon_0}\}\leq 
e^{-M^{\varepsilon_1}},\tag 3.10$$
provided $M$ is sufficiently large. 
\endproclaim
{\it Proof}.
  Let $e^{4i\Gamma_1^-(m)}\zeta^2(m)=
\xi_m(\rho)+i\eta_m(\rho)$, $m^{-2\b}w^2(\pi m)=
a_r(m)+ia_i(m)$,
$\xi_m,\eta_m, a_r(m), a_i(m) \in {\Bbb R}$. We will prove
(3.10) with $\Sigma_2(M)$ replaced by the sum
$$\sum\limits_{m=M_0}^M a_r(m)\xi_m(\rho)
\cos (4\Lambda^m\rho),\tag 3.11$$
the others cases being similar. 
We break  sum (3.11) into two parts:
$\sum_{M_0}^M=\sum_{M_0}^{M_1}+\sum_{M_1}^M$,
where $1\ll M_1 \ll M$ to be specified later.
The first sum can be estimated trivially
$$ \sum\limits_{m=M_0}^{M_1}a_r(m)\xi_m(\rho)
\cos (4\Lambda^m\rho) =O(M_1^{1-2\b}).\tag 3.12$$
Consider the second one. 
Let $L=\Lambda$, $h=M_1$, $f_n(\rho)=\xi_{M_1+n}(\Lambda^{-(M_1+n)}\rho)$.
Clearly, $\|f_n\|_\infty\leq 1$. By (2.59), (2.55), (2.56),  (3.6),
$$\left|{\partial\over \partial \rho}f_n(\rho)\right|
\leq C(M_1+n)^{-\b}.\tag 3.13$$
From (3.9), (3.13), 
one gets for any $\delta\geq 0$, $t\geq 0$, :
$$\mes \{ \rho \in I: \sum\limits_{m=M_1}^M a_r(m)\xi_m(\rho)
\cos (4\Lambda^m\rho)\geq \delta\}$$
$$\leq e^{-\delta t}\int_I d\rho
\exp\left(t\sum\limits_{m=M_1}^M a_r(m)\xi_m(\rho)
\cos (4\Lambda^m\rho)\right)$$
$$\leq e^{-\delta t}e^{Ct^2\sum_{M_1}^M m^{-4\b}+
Ct\sum_{M_1}^M m^{-3\b}},\tag 3.14 $$
provided $tM_1^{-2\b}$ is suficiently small:
$$q^2tM_1^{-2\b}\leq 1.$$ 

Choosing $t=c{\delta\over \sum_{M_1}^M m^{-4\b}}$ 
with a sufficiently small
constant $c$ one gets
$$\mes \{ \rho \in I: |\sum\limits_{m=M_1}^M a_r(m)\xi_m(\rho)
\cos (4\Lambda^m\rho)   |\geq \delta\}$$
$$\leq 2\exp\left(-C{\delta^2\over \sum_{M_1}^M m^{-4\b}}\right),
\tag 3.15$$
provided
$$c_1^{-1}\sum\limits_{m=M_1}^M m^{-3\b}\leq\delta
\leq c_1M_1^{2\b}\sum\limits_{m=M_1}^M m^{-4\b},\tag 3.16$$
for some suitable constant $c_1$.
We consider the cases
$$1. \quad 0< \b <  {1\over 3}$$
$$2.  \quad {1\over 3}\leq \b<{1\over 2}.$$
In the first  case we set
$$ \delta=M^{1-2\b-\varsigma_1},\quad M_1=M^{1-\varsigma_2},
\quad \varsigma_j>0.\tag 3.17$$
Choose $\varsigma_j$ in such a way that
$$ 0<\varsigma_1<\b,\quad 0<\varsigma_2<{\varsigma_1\over 2\b}.\tag 3.18$$ 
This gives
$$\delta^{-1}M^{1-3\b}, \delta M_1^{-2\b}
\left(\sum\limits_{m=M_1}^M m^{-4\b}\right)^{-1}=O(M^{-\g}),$$
which means that (3.16) is satisfied provided
$M$ is sufficiently large.
Combining (3.12), (3.15), (3.17) one gets (3.10) with,
say,
$$\varepsilon_0< {\b\over 2},\quad\varepsilon_1\leq 1-3\b.$$
\par Consider the case ${1\over 3}\leq \b<{1\over 2}$. Set
$$ \delta=M_1^{\varsigma_3},\quad M_1=M^{\varsigma_4},
\quad 0<\varsigma_4<1.$$
Then (3.16) is satisfied provided
$$1-3\b<\varsigma_3<1-2\b.$$
As a consequence, one obtains (3.10)
for any $\varepsilon_0$, $\varepsilon_1$ such that
$$\varepsilon_0<1-2\b,
\quad \varepsilon_1<1-2\b-\varepsilon_0.$$
\hfill$\blacksquare$
\par Since the right hand sides of (3.10) belong to $l_1$,
lemma 3.2 implies immediately
\proclaim{Lemma 3.3} For any $\a $ and a.e. $\rho$,
$$\lim\limits_{m\rightarrow+\infty}
{\Sigma_2(m)\over m^{1-2\b-\varepsilon}}=0,\tag 3.19$$
provided $\varepsilon<\varepsilon_0$.
\endproclaim
\subsubhead
3.1.3. Estimates of $\Sigma_1(M)$
\endsubsubhead
\par Consider  sum (3.7). Iteration of (3.3) gives
$$\zeta(m+T)=e^{2i\sum\limits_{s=1}^T\triangle \Gamma(m+T-s)}
\zeta(m)$$
$$+\bar r_0\sum\limits_{s=1}^T
(m+T-s)^{-\b}\overline{w(\pi (m+T-s))}
e^{2i\sum\limits_{k=1}^s\triangle \Gamma(m+T-k)-2i\Gamma_-(m+T-s)}$$
$$-r_0\sum\limits_{s=1}^T
(m+T-s)^{-\b}w(\pi (m+T-s))
e^{2i\sum\limits_{k=1}^s\triangle \Gamma(m+T-k)+2i\Gamma_-(m+T-s)}
\zeta^2(m+T-s)$$
$$+O(m^{-2\b}T),$$
for any $T>0$. 
Returning to (3.7) one gets for
$M$ sufficiently large and $T$,
$1\ll T\ll M$, to be specified below,
$$\Sigma_1(M)=\sum\limits_{m=M_0}^{2T} 
m^{-\b}w(\pi m)e^{2i\Gamma_-(m)}\zeta(m)$$
$$+\sum\limits_{m=2T}^{M} 
m^{-\b}w(\pi m)
e^{2i\Gamma_-(m)+2i\sum\limits_{s=1}^T\triangle \Gamma(m-s)}
\zeta(m-T)\tag 3.20$$
$$+\bar r_0\sum\limits_{s=1}^T\sum\limits_{m=2T}^{M}
m^{-2\b}w(\pi m)\overline{w(\pi(m-s))}
e^{2i\Gamma_-(m)-2i\Gamma_-(m-s)+
2i\sum\limits_{k=1}^T\triangle \Gamma(m-k)}\tag 3.21$$
$$-r_0\sum\limits_{s=1}^T\sum\limits_{m=2T}^{M}m^{-2\b}w(\pi m)
w(\pi(m-s))e^{2i\Gamma_-(m)+2i\Gamma_-(m-s)+
2i\sum\limits_{k=1}^T\triangle \Gamma(m-k)}\zeta^2(m-s)\tag3.22$$
$$+O(T\sum\limits_{m=2T}^M m^{-3\b})+O(T^{2-2\b}).$$
We choose $T$  as follows:
$$T=\left[{\ln M\over \ln\Lambda} \right]+1.$$ 
Then for
the first sum  one has the trivial estimate
$$\left|
\sum\limits_{m=M_0}^{M_1} 
m^{-\b}w(\pi m)e^{2i\Gamma_-(m)}\zeta(m)\right|\leq CT^{1-\b}
\leq C(\ln M)^{1-\b}.\tag 3.23$$
We next consider sum (3.20) and prove the followng 
 estimate.
\proclaim{Lemma 3.4} For any positive $z$ satisfying
$z<  1-2\b$, one has 
$$\mes\{\rho\in I: |(3.20)|\geq M^{1-2\b-z}\}\leq Ce^{-CM^{1-2\b-2z}},
\tag 3.24$$
provided $M$ is sufficiently large.
\endproclaim
\par {\it Proof}. The proof is similar to that of lemma 3.2.
Write $w(\pi m)m^{-\b}=a_r^1(m)+a_i^1(m)$,
$e^{2i\sum\limits_{s=1}^T\triangle \Gamma(m-s)}
\zeta(m-T)=\xi^1_m(\rho)+\eta^1_m(\rho)$,
$a^1_r(m), a^1_i(m), \xi^1(m), \eta^1(m)\in {\Bbb R}$.
It is sufficient to prove (3.24) for the sum
$$\sum\limits_{m=2T}^M a_r^1(m)\xi_m^1(\rho) \cos (2\Lambda^m\rho).$$
Clearly, one has,
$$|a^1(m)|\leq Cm^{-\b},\,\, |\xi^1_m(\rho)|\leq 1,\,\,
\left|{\partial\over \partial\rho} \xi^1_m(\rho)\right|\leq C
(\Lambda^{m-T}m^{-\b} +\Lambda_1^m).$$

Applying (3.9) one gets  for $t,\delta \geq 0$
$$\mes\{\rho\in I:|\sum\limits_{m=2T}^M a_r^1(m)\xi_m^1(\rho) 
\cos (2\Lambda^m\rho)|
\geq \delta\}$$
$$\leq 2\exp\left(-\delta t+Ct^2M^{1-2\b}+Ct\Lambda^{-T}M^{1-2\b}+
Ct\left({\Lambda_1\over \Lambda}\right)^{2T}\right),$$
provided
$$ CT^{-\b}t\leq 1.$$
This gives
$$\mes\{\rho\in I:|\sum\limits_{m=2T}^M a_r^1(m)\xi_m^1(\rho) 
\cos (2\Lambda^m\rho)|
\geq \delta\}$$
$$\leq 2e^{-C\delta^2M^{2\b-1}},$$
provided $\delta$ satisfies
$$ C(\Lambda^{-T}M^{1-2\b} +
\left({\Lambda_1\over \Lambda}\right)^{2T})\leq
\delta\leq C T^{\b}M^{1-2\b}.$$
Clearly, $\delta= M^{1-2\b-z}$
verifies these conditions. As a consequence, one gets (3.24).
\hfill $\blacksquare$
\par Notice that choosing 
$z<{1-2\b\over 2}$ one makes the r.h.s. of (3.24) 
to be in $l_1$, which implies:
\proclaim{Lemma 3.5} For a.e. $\rho$
$$\lim\limits_{m\rightarrow +\infty}{(3.20)\over m^{1-2\b-z}}=0,$$
provided $z<{1-2\b\over 2}$.
\endproclaim
Consider sums (3.21), (3.22).
\proclaim{Lemma 3.6} {\rm (3.21), (3.22)}
satisfy for some $\varepsilon_0, \varepsilon_1>0$,
$$\mes 
\{ \rho \in I: |(3.21)|+|(3.22)|\geq M^{1-2\beta-\varepsilon_0}\}\leq 
e^{-M^{\varepsilon_1}},\tag 3.25$$
provided $M$ is sufficiently large. 
\endproclaim

\par {\it Proof}.
Sums (3.21), (3.22) have the following structure
$$\sum\limits_{s=1}^T
\sum\limits_{m=2T}^{M} t(m,s)\Psi(m,s)e^{2i\phi(m,s)},$$ 
where
$$t(m,s)=\bar r_0m^{-2\b}w(\pi m)\overline{w(\pi(m-s))},\quad
\Psi(m,s)=e^{
2i\sum\limits_{k=1}^T\triangle \Gamma(m-k)},$$
$$\phi(m,s)=\Gamma_-(m)-\Gamma_-(m-s),$$
in the case of (3.21), and
$$t(m,s)= -r_0m^{-2\b}w(\pi m)w(\pi(m-s)),\quad
\Psi(m,s)=e^{2i\sum\limits_{k=1}^T\triangle \Gamma(m-k)}\zeta^2(m-s),$$
$$\phi(m,s)=\Gamma_-(m)+\Gamma_-(m-s),$$
 for (3.22). 

\par As in the proof of lemma 3.2 we write 
$t(m,s)=a_r(m,s)+ia_i(m,s)$, $\Psi(m,s)=\xi(m,s)+i\eta(m,s)$,
$a_r(m,s),\,a_i(m,s),\,\xi(m,s),\,\eta(m,s)\in {\Bbb R}$
 and
prove (3.25) for the sum
$$\sum\limits_{s=1}^T
\sum\limits_{m=2T}^{M}a_r(m,s)\xi(m,s)\cos(2\phi(m,s)).\tag 3.26$$
Clearly,
$$|a_r(m,s)|\leq C m^{-2\b},
\quad |\xi(m,s)|\leq 1,
\quad \left|{\partial\over \partial \rho}
\xi(m,s)\right|\leq C(\Lambda^{m-s}m^{-\b}+\Lambda_1^m),$$
$a_r(m,s)$ being independent of $\rho$.
Break sum (3.26) into two: $\sum_{2T}^M=\sum_{2T}^{M_1}+\sum_{M_1}^M$,
where $M_1, 2T\leq M_1\leq M$ to be choosen later.
For the first sum we have
$$\sum\limits_{s=1}^T
\sum\limits_{m=2T}^{M_1}a_r(m,s)\xi(m,s)\cos(2\phi(m,s))
=O(TM_1^{1-2\b}).$$
To estimate the second one we apply (3.9)
 with $L=\Lambda$, $k=1\pm \Lambda^{-s}$:
$$\mes\{\rho\in I:\sum\limits_{s=1}^T
\sum\limits_{m=M_1}^{M}a_r(m,s)\xi(m,s)\cos(2\phi(m,s))\geq \delta\}$$
$$\leq e^{-\delta t}\int_Id\rho
\exp\bigg(
t\sum\limits_{s=1}^T\sum\limits_{m=M_1}^{M}a_r(m,s)\xi(m,s)
\cos(2\phi(m,s)\bigg)$$
$$\leq e^{-\delta t} \prod\limits_{s=1}^T\left(\int_Id\rho
\exp\big(Tt\sum\limits_{m=M_1}^{M}a_r(m,s)\xi(m,s)\cos(2\phi(m,s)\big)
\right)^{1\over T}$$
$$\leq e^{-\delta t}
\exp\bigg(CT^2t^2\sum_{M_1}^Mm^{-4\b}+Ct\sum_{M_1}^Mm^{-3\b}\bigg),$$
provided
$$CTM_1^{-2\b}t\leq 1.$$
As a consequence, for $\delta$ satisfying 
$$C\sum_{M_1}^Mm^{-3\b}\leq \delta\leq CTM_1^{2\b}\sum_{M_1}^Mm^{-4\b},$$
one has
$$\mes\{\rho\in I:\sum\limits_{s=1}^T
\sum\limits_{m=M_1}^{M}a_r(m,s)\xi(m,s)\cos(2\phi(m,s))\geq \delta\}$$
$$\leq e^{-C{\delta^2\over T^2\sum_{M_1}^Mm^{-4\b}}}.$$
Therefore,  one can get the desired estimate (3.25)
by choosing $M_1$ and $\delta$ exactly in the same way as 
$M_1$ and $\delta$ in the proof of lemma 3.2.\hfill$\blacksquare$

\par Combining (3.23) and lemmas 3.5, 3.6, one gets
\proclaim{Lemma 3.7}
For any $\a$ and a.e. $\rho$,
$$\lim\limits_{m\rightarrow +\infty}
{\Sigma_1(m)\over m^{1-2\b-\varepsilon}}=0,$$
for some $\varepsilon >0.$
\endproclaim
This lemma together with lemma 3.3  lead to proposition 3.1.
 
\subhead
3.2. Decaying solutions of (3.1) 
\endsubhead
\par In this subsection we construct a decaying solution to (3.1).
Consider the solutions
$$\chi_0(m)=R_0(m){e^{i\varphi_0(m)}\choose e^{-i\varphi_0(m)}},\quad
\chi_{\pi/2}(m)=R_{\pi/2}(m)
{e^{i\varphi_{\pi/2}}(m)\choose e^{-i\varphi_{\pi/2}}(m)}.$$
One has
$$\det(\chi_0(m),\chi_{\pi/2}(m))=-2i,$$
which means that 
$$|\zeta_0(m)-\zeta_{\pi/2}(m)|={2\over R_0(m)R_{\pi/2}(m)},$$
where
$ \zeta_\a(m)=e^{2i\varphi_\a(m)}$, $\a=0,\pi/2$.
Applying the results of the previos subsection one gets
for a.e. $\rho$
$$|\zeta_0(m)-\zeta_{\pi/2}(m)|=e^{-2r_*m^{1-2\b}(1+o(1))},
\quad m\rightarrow +\infty.\tag 3.27$$
Set
$$v(m)=\ln\left({R_0(m)\over R_{\pi/2}(m)}\right).$$
By (3.2), (3.7) $v(m)$ satisfies
$$|v(m+1)-v(m)|\leq Cm^{-\b}|\zeta_0(m)-\zeta_{\pi/2}(m)|
\leq e^{-2r_*m^{1-2\b}(1+o(1))}.\tag 3.28$$
So,
$v(m)$ has a limit $v_\infty$ as $m\rightarrow +\infty$, and
$$ |v(m)-v_\infty|\leq  e^{-2r_*m^{1-2\b}(1+o(1))}.\tag 3.29$$
It folows from (3.1) that $z(m)=e^{i(\varphi_0(m)-\varphi_{\pi/2}(m))}$
satisfies
$$
{z(m+1)\over z(m)}=
{R_0(m)R_{\pi/2}(m+1)(s(d_m)+\bar r(d_m)e^{-2i\Gamma_-(m)}\zeta_0(m))
\over
R_0(m+1)R_{\pi/2}(m)
(s(d_m)+\bar r(d_m)e^{-2i\Gamma_-(m)}\zeta_{\pi/2}(m))}.$$

Combining this representation with (3.27), (3.28) one gets
$$|z(m+1)-z(m)|\leq e^{-2r_*m^{1-2\b}(1+o(1))}.$$
This means that $ z(m)$ has a limit
$z_\infty$ and
$$ |e^{i\varphi_0(m)}-z_\infty e^{i\varphi_{\pi/2}(m)}|
\leq e^{-2r_*m^{1-2\b}(1+o(1))}.\tag 3.30$$
Notice that by (3.27) $z_\infty^2=1$.
\par Proposition 3.1 and (3.29), (3.30) lead  immediately to 
the following result.
\proclaim{Proposition 3.2}
For a.e $\rho$ there exists
a real constant $h$ ($=-z_\infty e^{v_\infty}$) such that
the solution $\chi_0(m)+h\chi_{\pi/2}(m)$ satisfies
$$|\chi_0(m)+h\chi_{\pi/2}(m)|\leq e^{-r_*m^{1-2\b}(1+o(1))},$$
as $m\rightarrow \infty$.
\endproclaim
\head
4. End of the proof of  theorem 1.2
\endhead
\subhead 
4.1. Growing and decaying solutions of (2.60)
\endsubhead
\par Let us consider the solution $Q^\a$ of (2.1) corresponding to the 
following initial data:
$$ Q^\a(\xi(x))|_{x=x_{qk_M}}=
R_\a(M)\sin \bigg(\xi(x_{qk_M})+\varphi_\a(M)\bigg),$$
$$
Q^{\a}_\xi(\xi(x))|_{x=x_{qk_M}}=
R_\a(M)\cos \bigg(\xi(x_{qk_M})+\varphi_\a(M)\bigg),$$
where $M$ is sufficiently large number,
$\chi_\a(m)=R_\a(m){e^{i\varphi_\a(m)}\choose e^{-i\varphi_\a(m)}}$
is a solution of (3.1), 
$\chi_\a(M_0)={e^{i\a}\choose e^{-i\a}}$, $M_0$ being the same as in
section 3. We denote by $\hat\chi_\a(m)$ the $\hat\chi(m)$
corresponding to $Q^\a$:
$$\hat\chi_\a(m)=\tilde R(k_m)
{e^{i\tilde\varphi(k_m)}\choose e^{-i\tilde\varphi(k_m)}}=
R(x_{qk_m}){e^{i\varphi(x_{qk_m})}\choose e^{-i\varphi(x_{qk_m})}},$$
$R$, $\varphi$ being the   Pr\"ufer coordinates associated to $Q^\a$
$$ Q^\a=R\sin(\xi+\varphi),\quad Q^{\a}_\xi=R\cos(\xi+\varphi).$$
 One has the following proposition.
\proclaim{Proposition 4.1} For any $\a$ and a.e $\rho$ 
there exist real constants $g^1_\a\neq 0$, $g_\a^2$ such that
as $m\rightarrow +\infty$,
$$\hat\chi_\a(m)=g_\a^1\chi_\a(m) + g_\a^2\chi^d(m) +O(e^{-\g m}),$$
provided $M$ is chosen sufficiently large (it may depend on $\rho$).
Here $\chi^d(m)$ is the decaying solution of {\rm (3.1)} introduced in
proposition {\rm 3.1:} $\chi^d(m)=\chi_0(m)+h\chi_{\pi/2}(m)$.
\endproclaim
\par {\it Proof}. $\hat\chi_\a(m)$
 satisfies
$$\hat\chi_\a(m+1)=\A_0(m)\hat\chi_\a(m)+{\Cal R}(m),
\quad \hat\chi_\a(M)=\chi_\a(M),\tag 4.1$$
where
$$|{\Cal R}(m)|\leq Ce^{-\g m}|\hat\chi_\a(m)|.$$
We apply to (4.1) the following variation parameter type transformation:
$$\hat\chi_\a(m)={\bold \Psi}(m)g(m),
\quad {\bold \Psi}(m)=(\chi_\a(m), \chi^d(m)),$$
$\det {\Bbb \Psi(m)}\neq 0$ for a.e. $\rho$.
This transformation brings (4.1) to the form
$$ g(m+1)=g(m)+ \tilde{\Cal R}(m),\quad 
\tilde {\Cal R}(m)={\bold \Psi}^{-1}(m){\Cal R}(m),
\quad g(M)={1\choose 0}.\tag 4.2$$
Clearly, 
$$|\tilde{\Cal R}(m)|\leq Ce^{-\g m}|g(m)|.$$
One can rewrites (4.2) in the form
$$g(m+1)={1\choose 0} +\sum\limits_{j=M}^m \tilde {\Cal R}(j),
\quad m\geq M.$$
As a consequence, for $M$ sufficiently large, one has
$$|g(m)-{1\choose 0}|\leq C e^{-\gamma M},\quad m\geq M,$$
which, in particular, implies that
$$|g(m+1)-g(m)|\leq Ce^{-\g m}.$$
So, as $m\rightarrow +\infty$, $g(m)$ has a limit 
$g_\infty={g_\a^1\choose g_\a^2}$ and
$$|g(m)-g_\infty|\leq Ce^{-\g m},$$
$$|g_\a^1-1|,\,|g_\a^2|\leq Ce^{-\g M}.$$
Returning to $\hat \chi_\a$ one gets
$$\hat\chi_\a(m)={\bold \Psi}(m)g_\infty +O(e^{-\g m}).$$
Note that $\overline{ g(m)}=g(m)$, so,  $g^j_\a$ are real.
\hfill  $\blacksquare$

\par The decaying solution of (2.60) can be now constructed as follows.
Consider the solution $Q^d$ of (2.2) defined by
$$Q^d=
 Q^0+h_1Q^{\pi/2},$$
where $h_1={hg_0^1\over g_{\pi/2}^1}\in {\Bbb R}$.
Then the corresponding $\hat\chi(m)$ has the form
$$\hat\chi(m)=\hat\chi_0(m)+h_1\hat\chi_{\pi/2}(m),$$
and by propositions  4.1, admits the estimate
$$\hat\chi(m)= h_2\chi^d(m)+O(e^{-\g m}),$$
for some constant $h_2$.
In particular,
$$|\hat\chi(m)|\leq e^{-r_*m^{1-2\b}(1+o(1))}.\tag 4.3$$
Notice also that
$$\hat\chi(M)=\chi_0(M)+h\chi_{\pi/2}(M)\neq 0,$$
which means that $Q^d$ is nontrivial solution.
\par We are now able to complete the proof of theorem 1.2.
Let $R_d$ and $R_i$ be $R$'s associated to $Q^d$ and $Q^0$ respectively.
Combining  propositions 3.1, 3.2, 4.1 and (4.3),
(2.52), (2.51), (2.30), (2.13), (2.7)
 one gets the following result.
\proclaim{Proposition 4.2} For a.e.$\rho$, $R_d$, $R_i$ satisfy
$$\int_0^N dx p^{-1}R_d^2\leq N^{1/2}e^{-\mu_*(\ln N)^{1-2\b}(1+o(1))},$$
$$\int_0^N dx p^{-1}R_i^2\geq N^{1/2}e^{\mu_*(\ln N)^{1-2\b}(1+o(1))},$$
provided $N$ is sufficiently large. 
Here $\mu_*={r_*\over (2\ln \Lambda)^{1-2\b}}$.
\endproclaim
 This means in particular that for a.e $\rho$, $Q^d$ is a subordinate 
solution of (2.2), which completes the proof of theorem 1.2. 
\head
Appendix 1
\endhead
In this appendix we prove lemmas 2.1, 2.2, 2.3.

\par {\it Proof of lemma} 2.1. 
First we consider the integral
$$J=\int_{x_l}^x  dy \psi_l(y)e^{2i\xi(y)}
v_l(y),$$
where
$\psi_l(x)$ stands for either
$\chi_l(x)$ or $1-\chi_l(x)$,
$x_l\leq x\leq x_{l+1}$,
$v_l(x)=v(x)-\hat v(l) e^{-2i\pi l x}$,
$v\in L_1({\Bbb T})$, $\chi_l$ being described in lemma 2.1. 
It admits the estimate
$$|J|\leq 
c\|v\|_{L_1({\Bbb T})}.\tag A1.1$$
Indeed, for $|x-x_l|\leq 4$, (A1.1) is trivial.
For $|x-x_l|\geq 4$ we write the integral
$J$ as the sum
$$ J=J^{(0)}+J^{(1)},$$
$$J^{(0)}=\int_{x_l}^x  dy e^{2i\xi(y)}(\chi(y-x_l)+
\chi(y-x)\psi_l(y)v_l(y),$$
$$J^{(1)}=\int_{x_l}^x  dy e^{2i\xi(y)}(1-\chi(y-x_l)-
\chi(y-x))\psi_l(y)v_l(y).$$

Clearly,
$$|J^{(0)}|\leq c\|v_l\|_{L_1({\Bbb T})}
\leq c\|v\|_{L_1({\Bbb T})}.\tag A1.2$$
To estimate $J^{(1)}$
we represent it as the sum
$$J^{(1)}=\sum\limits_{n,n\neq l}\hat v_n \zeta_n,$$
$$\zeta_n=\int_{x_l}^x  dy (1-\chi(y-x_l)-
\chi(y-x))\psi_l(y)e^{2i(\xi(y)-\pi ny)}.$$
Integrating by parts one gets the representation
$$\zeta_n=-{1\over 4}\int\limits_{x_l}^x  
dy e^{2i(\xi(y)-\pi ny)}\!
\left(\!{d\over dy}{1\over (p(y)-\pi n )}\!\right)^2\!(1-\chi(y-x_l)-
\chi(y-x))\psi_l(y),$$
$$p(y)=(Fy+q(y)+E)^{1/2},$$
which leads to the estimate
$$|\zeta_n|\leq c<n-l>^{-2}, n\neq l.$$
As a consequence,
$$|J^{(1)}|\leq c\|v\|_{L_1({\Bbb T})}.\tag A1.3$$
Combining (A1.2, A1.3), we get (A1.1).
\par Consider the integrals
$$J_1=\int_{x_l}^x dy e^{2i(\xi(y)- \pi ly)}(1-\chi_l(y)),$$
$$J_2=\int_{x_l}^x dy e^{2i(\xi(y)- \pi ly)}\chi_l(y).$$
Since for $|x-X_l|\geq c l^{1-\nu}$,
$$|\xi(x)-\pi l|\geq l^{-\nu},$$
the integration  by parts in the first one gives  immediatly
$$ |J_1|\leq cl^{\nu}.\tag A1.4$$
The second integral can be represented in the form
$$J_2=e^{2i(\xi(X_l)-\pi l X_l)}
\int_{x_l}^x dy e^{i\mu_1(l)(y-X_l)^2+i\mu_2(l)(y-X_l)^3}
\chi_l(y) +O(l^{-3\nu}),\tag A1.5$$
where
$$\mu_1(l)={F\over 2\pi l},\quad \mu_2(l)=-{F^2\over 12(\pi l)^3}.$$
(A1.5) implies directly
$$ |J_2|\leq cl^{1/2}.\tag A1.6$$
Combining (A1.1), (A1.4), (A1.6) one obtains lemma 2.1.
\hfill  $\blacksquare$
\par {\it Proof of lemma 2.2}.
First  we remark that up to the terms of order
$O(l^{1/2})$ the expression
$I_l(E)$ can be replaced by $I_l^*(E)$,
$$I_l^*(E)=\int\limits_{x_l^*}^{x_{l+1}^*}dy
\int\limits_y^{x_{l+1}^*}dse^{2i(\xi(s)-\xi(y))}
v(s)\overline{v(y)},$$
where $x_l^*$ defined by
$$ Fx_l^* +q(x_l^* )=\pi^2(l-1/2)^2.$$
Indeed, one has 
$x_l^*-x_l=O(1)$, which together with lemma 2.1 implies
$$I_l(E)=I_l^*(E) + O(l^{1/2}).\tag A1.7$$
\par To estimate  $I_l^*(E)$ we write it as the sum
$$I_l^*(E)={\Cal I}_l^0(E)+{\Cal I}_l^1(E)+{\Cal I}_l^2(E)
+{\Cal I}_l^3(E),\tag A1.8$$
$${\Cal I}_l^0(E)=|\hat v(l)|^2\int_{x_l^*}^{x_{l+1}^*}dy
\int_y^{x_{l+1}^*}ds e^{2i(\xi(s)-\xi(y)-\pi  l(s-y))}$$
$${\Cal I}_l^1(E)=
\int_{x_l^*}^{x_{l+1}^*}dy\int_y^{x_{l+1}^*}dse^{2i(\xi(s)-\xi(y))}
v_l(s)\overline{v_l(y)}$$
$${\Cal I}_l^2(E)=\overline{\hat v(l)}
\int_{x_l^*}^{x_{l+1}^*}dy
\int_y^{x_{l+1}^*}dse^{2i(\xi(s)-\xi(y)+\pi  ly)}
v_l(s)$$
$${\Cal I}_l^3(E)=\hat v(l)
\int_{x_l^*}^{x_{l+1}^*}dy
\int_y^{x_{l+1}^*}dse^{2i(\xi(s)-\xi(y)-\pi  ls)}
\overline{v_l(y)}$$
$$=\hat v(l)
\int_{x_l^*}^{x_{l+1}^*}dy\int^y_{x_{l}^*}
dse^{2i(\xi(y)-\xi(s)-\pi  ly)}
\overline{v_l(s)}.\tag A1.9$$
It follows from (A1.1)
that for $x_l^*\leq y\leq x_{l+1}^*$,
$$|\int_y^{x_{l+1}^*}dse^{2i\xi(s)}
v_l(s)|\leq c\|v\|_1.$$
As a consequence,
$$|{\Cal I}_l^j(E)|\leq C l,\quad j=1,2,3.\tag A1.10$$
Consider the derivatives ${d{\Cal I}_l^j\over dE}$, $j=1,2$.
We write them as
$${d{\Cal I}_l^1\over dE}(E)=
i\int\limits_{x_l^*}^{x_{l+1}^*}\!\!dy\int\limits_y^{x_{l+1}^*}
\!\!dse^{2i(\xi(s)-\xi(y))}
\left(\int\limits_{x_l^*}^s \!d\rho p^{-1}(\rho, E)-
\int\limits_{x_l^*}^y\!d\rho p^{-1}(\rho, E)\right)
v_l(s)\overline{v_l(y)},$$
$${d{\Cal I}_l^2\over dE}(E)=i\overline{\hat v(l)}
\int\limits_{x_l^*}^{x_{l+1}^*}\!\!dy\int\limits_y^{x_{l+1}^*}
\!dse^{2i(\xi(s)-\xi(y)+\pi  ly)}
\left(\int\limits_{x_l^*}^s \!d\rho p^{-1}(\rho, E)\right.$$
$$-
\left.\int\limits_{x_l^*}^y\!d\rho p^{-1}(\rho, E)\right)
v_l(s).$$
Clearly,
 for $x_l^*\leq y\leq x_{l+1}^*$ one has
$$\left|\int_{x_l^*}^yd\rho p^{-1}(\rho, E)\right|\leq  c,$$
$$|\int_y^{x_{l+1}^*}dse^{2i\xi(s)}\int_{x_l^*}^sd\rho p^{-1}(\rho, E)
v_l(s)|\leq c\|v\|_1.$$
As a consequence,
$$\left|{d{\Cal I}_l^j\over dE}\right|
\leq Cl, \quad j=1,2.\tag A1.11$$
By (A1.9),
the same estimate is valid for 
 ${d{\Cal I}_l^3\over dE}$
$$\left|{d{\Cal I}_l^3\over dE}\right|
\leq Cl.\tag A1.12$$
\par To analyse the expression ${\Cal I}_l^0$ we represent it as
$${\Cal I}_l^0 =|\hat v(l)|^2({\Cal I}_l^{00} +{\Cal I}_l^{01}+
{\Cal I}_l^{02}), $$
$${\Cal I}_l^{00}=\int_{x_l^*}^{x_{l+1}^*}dy\tilde\chi_l(y)
\int_y^{x_{l+1}^*}ds \chi_l(s)e^{2i(\xi(s)-\xi(y)-\pi  l(s-y))},$$
$${\Cal I}_l^{01}=\int_{x_l^*}^{x_{l+1}^*}dy\tilde\chi_l(y)
\int_y^{x_{l+1}^*}ds(1-\chi_l(s))e^{2i(\xi(s)-\xi(y)-\pi  l(s-y))},$$
$${\Cal I}_l^{02}=\int_{x_l^*}^{x_{l+1}^*}dy(1-\tilde\chi_l(y))
\int_y^{x_{l+1}^*}dse^{2i(\xi(s)-\xi(y)-\pi  l(s-y))}.$$
Here $\chi_l(x)=\chi(l^{-1+\nu}(x-X_l))$,
$\tilde\chi_l(x)=\chi(l^{-1+\tilde\nu}(x-X_l))$,
$0<\nu<\tilde\nu<1/2$.
\par First we consider ${\Cal I}_l^{02}$. Integrating by parts
and using (A1.4), (A1.6) one gets
$$
{\Cal I}_l^{02}={i\over \pi}
\int_{x_l^*}^{x_{l+1}^*} ds e^{2i(\xi(s)-\pi ls)}
+O(l^{-1/2})
$$
$$
-{i\over 2}\int_{x_l^*}^{x_{l+1}^*} dy e^{-2i(\xi(y)-\pi ly)}
{d\over dy}\left({1\over p(y) -\pi l}(1-\tilde\chi_l(y))
\int_y^{x_{l+1}^*} ds e^{2i(\xi(s)-\pi ls)}\right)
$$ 
$$=O(l^{1/2+\tilde\nu})+{i\over 2}\int_{x_l^*}^{x_{l+1}^*} dy
{1-\tilde\chi_l(y)\over p(y) -\pi l}.$$
Consider the integral
$$\int_{x_l^*}^{x_{l+1}^*} dy
{1-\tilde\chi_l(y)\over p(y) -\pi l}.
 \tag A1.13$$
One has
$$(A1.13)=
\left(\int_{x_l^*}^{X_{l}-2l^{1-\tilde\nu}} dy
+\int^{x_{l+1}^*}_{X_{l}+2l^{1-\tilde\nu}} dy
\right){1\over p(y)-\pi l}$$
$$+{2\pi l\over F+q^\prime(X_l)}
\int\limits_{|y-X_l|\leq 2l^{1-\tilde\nu}} dy
{1-\tilde\chi_l(y)\over y-X_l}+O(l^{-\tilde\nu}).$$
Since $\chi$ is an even function
the last integral here vanishs. Therefore, one has
$$(A1.13)=
\left(\int_{x_l^*}^{X_{l}-2l^{1-\tilde\nu}} dy
+\int^{x_{l+1}^*}_{X_{l}+2l^{1-\tilde\nu}} dy
\right){1\over p(y)-\pi l}+O(l^{-\tilde\nu})$$
$$={2\over F}\left(\int_{x_l^*}^{X_{l}-2l^{1-\tilde\nu}} dy
+\int^{x_{l+1}^*}_{X_{l}+2l^{1-\tilde\nu}} dy
\right){p^\prime p\over p-\pi l}+O(1)
=O(1).$$
As a consequence,
$$|{\Cal I}_l^{02}|\leq C l^{1-\g}.\tag A1.14$$
\par Consider ${\Cal I}_l^{01}$. 
By (A1.4),
 $$|{\Cal I}_l^{01}|\leq C l^{1+\nu-\tilde\nu}.\tag A1.15$$
\par The expression  ${\Cal I}_l^{00}$
admits the representation
$$ {\Cal I}_l^{00}=\int\limits_{x_l^*}^{x_{l+1}^*}dy\tilde\chi_l(y)
\int\limits_y^{x_{l+1}^*}ds 
\chi_l(s)e^{i\mu_1(l)(X_l)((s-X_l)^2-(y-X_l)^2)}
+O(l^{2-4\nu-\tilde\nu})$$
$$=\int\limits_{-\infty}^\infty dy\chi(l^{-1+\tilde\nu}y)
\int_y^\infty ds
\chi(l^{-1+\nu}s)e^{i\mu_1(l)(s^2-y^2)}
+O(l^{2-4\nu-\tilde\nu})$$
$$=2\int_0^\infty dy \chi(l^{-1+\tilde\nu}y)e^{-i\mu_1(l)y^2}
\int_0^\infty ds
\chi(l^{-1+\nu}s)e^{i\mu_1(l)s^2}
+O(l^{2-4\nu-\tilde \nu}).\tag A1.16$$
At the last step here we used the fact that $\chi $ is an even function.
The expression (A1.16) allows some further simplifications
$$(A1.16)={\pi^2l\over F} +O(l^{1/2+\tilde\nu})+
+O(l^{2-4\nu-\tilde\nu}).$$
Choosing 
$${1\over 5}<\nu<\tilde\nu<{1\over 2},$$
one gets 
$$  {\Cal I}_l^{00}={\pi^2 l\over F}+O(l^{1-\g}),\tag A1.17$$
or combining (A1.14), (A1.15), (A1.16)
$$  {\Cal I}_l^{0}=|\hat v(l)|^2{\pi^2 l\over F}+O(l^{1-\g}),\tag A1.18$$
which together with (A1.7), (A1.8), (A1.10), (A1.11), (A1.18) 
gives the representation of lemma 2.2,
${\Cal I}_l(E)$ being given by
$${\Cal I}_l(E)=
|\hat v(l)|^2{\pi^2l\over F}+{\Cal I}_l^1(E)+{\Cal I}_l^2(E)
+{\Cal I}_l^3(E).\tag A1.19$$
\hfill  $\blacksquare$
\par {\it Proof of lemma 2.3}.
Consider the integral
$$
\int_{x_l}^{x_{l+1}}  dye^{2i\xi(y)}
v(y).\tag A1.20$$
We break it into three parts
$$\int_{x_l}^{x_{l+1}}dy=\int_{x_l}^{L^-_{l}+1/2}dy+
\int_{L_l^+-1/2}^{x_{l+1}}dy +\int^{L_l^+-1/2}_{L^-_{l}+1/2}dy,$$
where $L_l^\pm=[X_l\pm l^{1-\nu}]$, $0<\nu<1/2$ to be fixed later.
We 
start by estimating 
the two first integrals here. Write them  as the sums
$$\int_{x_l}^{L^-_{l}+1/2}dye^{2i\xi(y)}
v(y)+
\int_{L_l^+-1/2}^{x_{l+1}}dye^{2i\xi(y)}
v(y)$$
$$=\bigg(\sum\limits_{n=x_l+1/2}^{L_l^-}
+\sum\limits^{x_{l+1}-1/2}_{n=L_l^+ }\bigg)
\int\limits_{n-1/2}^{n+1/2}dye^{2i\xi(y)}
v(y).\tag A1.21$$
For $|n-X_l|\geq Cl^{1-\nu}$, the expression
$\int\limits_{n-1/2}^{n+1/2}dye^{2i\xi(y)}
v(y)$ has the form
$$\int\limits_{n-1/2}^{n+1/2}dye^{2i\xi(y)}v(y)=
e^{2i\xi(n)}\int_{-1/2}^{1/2}dye^{2i\xi^\prime(n)y}
(1+i\xi^{\prime\prime}(n)y^2)v(y) +O(l^{-2}).\tag A1.22$$
Under assumptions (2.12) one has for any integer $m$, $m\geq 1$,
$k\in {\Bbb R}$,
$$\int_{-1/2}^{1/2} dy e^{2iky}y^mv(y)=O(k^{-1}).\tag A1.23$$
So, the r.h.s. of (A1.22) can be simplified:
$$\int\limits_{n-1/2}^{n+1/2}dye^{2i\xi(y)}v(y)
=e^{2i\xi(n)}V(n)+O(l^{-2}),\quad V(n)=
\hat v\bigg({\xi^\prime(n)\over \pi}\bigg),$$
$\hat v(k)=\int_{-1/2}^{1/2} dy e^{2i k\pi y}v(y)$.
As a consequence, one obtains
$$(A1.21)=\bigg(\sum\limits_{n=x_l+1/2}^{L_l^-}
+\sum\limits^{x_{l+1}-1/2}_{n=L_l^+ }\bigg)
e^{2i\xi(n)}V(n) +O(l^{-1})$$
$$=
e^{2i\xi(L_l^-)}{V(L_l^- )\over 
1-e^{-2i\xi^{(1)}(L_l^-)}}-e^{2i\xi(L_l^+-1)}{V(L_l^+)\over 
1-e^{-2i\xi^{(1)}(L_l^+)}}
\tag A1.24$$
$$
+e^{2i\xi(x_{l+1}-1/2)}{V(x_{l+1}-1/2)\over
1-e^{-2i\xi^{(1)}(x_{l+1}-1/2)}}-
e^{2i\xi(x_{l}-1/2)}{V(x_{l}+1/2)\over
1-e^{-2i\xi^{(1)}(x_{l}+1/2)}}\tag A1.25$$
$$+\bigg(\sum\limits_{n=x_l+1/2}^{L_l^--1}
+\sum\limits^{x_{l+1}-{3\over 2}}_{n=L_l^+ }\bigg)
e^{2i\xi(n)}V_1(n) +O(l^{-1}).\tag A1.26$$
Here
$$\xi^{(1)}(n)=\xi(n)-\xi(n-1),\quad
V_1(n)={V(n)\over 1-e^{-2i\xi^{(1)}(n)}}-
{V(n+1)\over 1-e^{-2i\xi^{(1)}(n+1)}}.$$
Clearly, $V_1(n)$ satisfies
$$ |V_1(n)|\leq Cl^{-1+2\nu},$$
$$\left |{V_1(n)\over 1-e^{-2i\xi^{(1)}(n)}}-
{V_1(n+1)\over 1-e^{-2i\xi^{(1)}(n+1)}}\right|\leq C
l^{-2+2\nu}(\sin(p(n))^{-2}.$$
So, repeating the procedure one gets 
$$\bigg(\sum\limits_{n=x_l+1/2}^{L_l^--1}
+\sum\limits^{x_{l+1}-3/2}_{n=L_l^+}\bigg)
e^{2i\xi(n)}V_1(n)=O(l^{-1+3\nu}).\tag A1.27$$
Consider the expressions (A1.24), (A1.25).
The first one may be represented
as
$$(A1.24)=i{\pi\over F}\hat v_ll^{\nu}
\big(e^{i\xi(L_l^+-1)+{F\over 2\pi}l^{-\nu}}
+e^{i\xi(L_l^-)-{F\over 2\pi}l^{-\nu}}\big) +O(l^{-\gamma}).\tag A1.28$$
Expression (A1.25) has the form
$$(A1.25)=t_{l+1}-t_{l} +O(l^{-1}),\tag A1.29$$
where 
$$ t_l=e^{2i\xi(x_{l}-1/2)}{V(x_{l}-1/2)\over
1-e^{-2i\xi^{(1)}(x_{l}-1/2)}}.$$
Clearly,
$$|t_l|\leq C.$$ 
Combining (A1.27), (A1.28), (A1.29),
one obtains
$$(A1.21)=i{\pi\over F}\hat v_ll^{\nu}
\big(e^{i\xi(L_l^+-1)+{F\over 2\pi}l^{-\nu}}
+e^{i\xi(L_l^-)-{F\over 2\pi}l^{-\nu}}\big)$$
$$+t_{l+1}-t_{l}+O(l^{-1+3\nu})+O(l^{-\gamma}).
\tag A1.30$$

Next we consider the expression
$$\int^{L_l^+-1/2}_{L_l^-+1/2}dye^{2i\xi(y)}
v(y)=\sum\limits_{n=L_l^-+1}^{L_l^+-1}
\int\limits_{n-1/2}^{n+1/2}dye^{2i\xi(y)}v(y).$$
For $|n-X_l|\leq Cl^{1-\nu}$, one has
$$\int\limits_{n-1/2}^{n+1/2}dye^{2i\xi(y)}v(y)=
e^{2i\xi(X_l)-2i\pi l X_l}e^{i\mu_1(l)(n-X_l)^2}
\hat v_l
+O(l^{-3\nu}).$$
We fix now $\nu$ in such a way that
$$1/4<\nu<1/3.$$
As a consequence, one obtains
$$\int^{L_l^+-1/2}_{L_l^-+1/2}dye^{2i\xi(y)}
v(y)=e^{2i\xi(X_l)-2i\pi l X_l}\sum\limits_{n=L_l^-+1}^{L_l^+-1}
e^{i\mu_1(l)(X_l)(n-X_l)^2}
\hat v_l
+O(l^{-\g})$$
$$
=e^{2i\xi(X_l)-2i\pi l X_l}\hat v_l\int^{L_l^+}_{L_l^-+1}dy
e^{i\mu_1(l)(y-X_l)^2} +O(l^{-\g}).\tag A1.31$$

The integral in the r.h.s may be represented as
$$ e^{2i\xi(X_l)-2i\pi l X_l}\hat v_l\int^{L_l^+}_{L_l^-+1}dy
e^{i\xi^{\prime\prime}(X_l)(y-X_l)^2}=e^{2i\omega_l}\pi 
\left({2l\over F}\right)^{1/2}$$
$$
-i{\pi\over F}\hat v_ll^{\nu}\bigg(e^{2i\xi(L^+_l)}+
e^{2i\xi(L^-_l+1)}\bigg)+O(l^{-\g}).\tag A1.32$$
Combining (A1.30), (A1.31), (A1.32) one gets lemma 2.3.
\hfill  $\blacksquare$

\head
Appendix 2
\endhead
Here we prove lemma 3.1.  
The proof is based on the nequality:
$$e^z\leq 1+z +Bz^2,\tag A2.1$$
valid, say, for $|z|\leq 1$.
By (A2.1),
$$\int_I dy\exp(tS_N(y))\leq X_N(t),\quad
X_N(t)=\int_I dy {\Cal X}_N(y,t),\tag A2.2$$
$$   {\Cal X}_N(y,t)=
\prod\limits_{n=0}^N (1+ts_n(y)+Bt^2s_n^2(y)),$$ 
$$s_n(y)=a(n)f_{n}(L^{h+n}y)g(L^{h+n}y),$$
provided
$$ t\max\limits_{0\leq n\leq N}|a(n)|\leq 1.
\tag A2.3$$
We represent $X_N(t)$ as follows
$$
X_N(t)=X_{N-1}(t)+ ({\roman I})+(\roman {II}),$$
$$(\roman{I})=Bt^2a^2(N)\int_I dy 
 {\Cal X}_{N-1}(y,t)
f^2_{N}(L^{h+N}y)g^2(L^{h+R}y)$$
$$(\roman {II})=
ta(N)C(t),\quad
C(t)=
\int_I dy  {\Cal X}_{N-1}(y,t)f_{N}(L^{h+N}y)g(L^{h+N}y).
$$

For $(\roman{I})$ one has the obvious estimate
$$|(\roman{I})|\leq Bt^2a^2(N)X_{N-1}(t).\tag A2.4$$
Consider expression $(\roman{II})$.
One can write $C(t)$ in the form
$$C(t)=C^1(t)+C^2(t),$$
$$C^1(t)=
\int_I dy  {\Cal X}_{N-1}(y,t) f_{N}(L^{h+N}y)$$
$$\times
\left(g(L^{h+N}y)-\int_Id\xi g(L^{h+N}\xi)\right) $$
$$C^2(t)=\left(\int_Id\xi g(L^{h+N}\xi)\right)
\left(\int_I dy  {\Cal X}_{N-1}(y,t)f_{N}(L^{h+N}y)\right).$$
Let
$$g(L^{h+N}y)-\int_Id\xi g(L^{h+N}\xi)=h^\prime(y),$$
where $h=0$ at the end points of $I$.
Clearly,
$$ \left|\int_Id\xi g(L^{h+N}\xi)\right|\leq 2KL^{-h-N},$$
$$ |h(y)|\leq 4KL^{-h-N},\quad y\in I.$$
Therefore,
$$|C^2(t)|\leq  2KL^{-h-N}X_{N-1}(t),\tag A2.5$$
$$|C^1(t)|\leq B(K,\Lambda)X_{N-1}(t)\left(\|f_N^\prime\|_\infty
+ L^{-h-N}t\sum\limits_{j=0}^{N-1}\|s_j^\prime\|_\infty\right)
$$
$$\leq B(K,\Lambda)X_{N-1}(t)\left(\|f_N^\prime\|_\infty+
t\sum\limits_{j=0}^{N-1}|a(j)|L^{j-N}
\right).\tag A2.6$$
Combining (A2.5), (A2.6) one gets
$$ |(\roman{II})|\leq B(K,\Lambda) X_{N-1}(t)t|a(N)|$$
$$\times\left(L^{-h-N}+
\|f_N^\prime\|_\infty+
t\sum\limits_{j=0}^{N-1}|a(j)|L^{j-N}
\right),\tag A2.7$$
which together with (A2.4) leads to the inequality
$$X_{N}(t)\leq X_{N-1}(t)\bigg(1+Bt^2a^2(N)$$
$$+
B(K,\Lambda)t|a(N)|\big(L^{-h-N}+\|f_N^\prime\|_\infty+
t\sum\limits_{j=0}^{N-1}|a(j)|
L^{j-N}\big)\bigg).\tag A2.8$$
In the same way one obtains
$$X_{0}(t)\leq 1+Bt^2a^2(0)+B(K,\Lambda)
t|a(0)|(L^{-h}+\|f_0^\prime\|_\infty).\tag A2.9$$
Combining (A2.8), (A2.9) one gets
$$X_N(t)\leq \exp\bigg(Bt^2A^2(N)+B(K,\Lambda)\big(t
{\Cal Q}(N)+
t^2\sum\limits_{n=1}^N\sum\limits_{j=0}^{n-1}
|a(n)||a(j)|L^{j-n}\big)\bigg) $$
$$\leq \exp\bigg(B(K,\Lambda)t^2A^2(N)+B(K,\Lambda)t{\Cal Q}(N)\bigg),$$
where 
$$A^2(N)=\sum\limits_{n=1}^Na^2(n),\quad
{\Cal Q}(N)=\sum\limits_{n=0}^N|a(n)|(L^{-h-n}+
\|f_n^\prime\|_\infty).$$
\hfill $\blacksquare$

\head
Appendix 3
\endhead
In this appendix we prove the following proposition.

\proclaim{Proposition A3.1}
Let $v\in L_1({\Bbb T})$. Then
the essential spectrum
of operator {\rm (1.1)},  $\s_{ess}(H)$, fills up the real axis:
$$\s_{ess}(H)={\Bbb R}.\tag A3.1$$
\endproclaim
{\it Proof}. 
We prove (A3.1) by employing the constructions of [16].
We start by replacing the whole line operator by a half-line operator
by putting a Dirichlet condition at $x=R$. This is a relative 
trace class
perturbation and it decomposes the whole-line operator into the direct sum 
of two half-line operators. Since
the spectrum of the left half-line operator is
discrete, one has
 $$ \sigma_{ess}(H)=\sigma_{ess}(H_R),$$
for any $R\in{\Bbb R}\,$.
Here $H_R$ denote right half-line operator with the Dirichlet condition
as $x=R$
$$ (H_R\psi)(x)=-\psi^{\prime\prime}-(Fx+q(x)-v(x))\psi(x),\quad x\geq R,
\quad \psi(R)=0.$$
So, to prove (A3.1) it is sufficient to show that
for some $R\in {\Bbb R}$ and $E\in {\Bbb C}$,
$\im E>0$, the difference
$$R_v(E)-R_0(E)\tag A3.2$$
is a compact operator.
Here $R_v(E)=(H_R-E)^{-1}$, $R_0(E)=(H_0-E)^{-1}$,
$H_0$ stands for the free
 right half-line operator with the Dirichlet condition
as $x=R$:
$$ (H_0\psi)(x)=-\psi^{\prime\prime}-(q(x)+Fx)\psi(x),
\quad x\geq R,
\quad \psi(R)=0$$
\par To calculate (A3.2) we write  the equation
$$
 (H_R-E)\psi =f,\quad \im E>0$$
 as a first-order system
$${\psi\choose\psi^\prime}^\prime=\pmatrix
0&1\cr
v-q-Fx-E&
0 \cr\endpmatrix 
{\psi\choose\psi^\prime}-{0\choose f}\tag A3.3$$
and
apply a variation of parameter-type transformation that brings the
system into nearly diagonalized form:
$${\psi\choose\psi^\prime}={\Bbb E}(x,E)\vec z,\quad {\Bbb E}(x,E)=
\pmatrix
1&1\cr
{\Cal E}(x,E)&{\Cal E}^*(x,E)\cr \endpmatrix ,\,\,
\vec z={z_1\choose z_2},$$
$${\Cal E}= {f_{as}^\prime\over f_{as}},\quad
{\Cal E}^*(x,E)= {f_{as}^{*\prime}\over f_{as}^{*}},\tag A3.4$$
where $f_{as}(x,E)$, $f^*_{as}(x,E)$
are standard WKB asymptotics corresponding to the 
equation
$-\psi_{xx}-(q(x)+Fx)\psi=E\psi$:
$$f_{as}(x,E)=p^{-1/2}(x,E)e^{i\Phi_{as}(x,E)},\quad
f_{as}^{*}(x,E)=p^{-1/2}(x,E)e^{-i\Phi_{as}(x,E)},$$
$$
p(x,E)= (Fx-q+E)^{1/2},\quad
\Phi_{as}(x,E)=\int_{R}^x dsp(s,E),$$
 $R$ is supposed to be sufficiently large.
The roots here  are defined on the complex plane with the cut along 
negative imaginary semi-axes and  is positive for positive values
of the arguments.

Applying (A3.4) to (A3.3)
we arrive at
$$ {\Cal H}_v(E)\vec z=\vec g, \quad 
\vec g={i\over 2}{f\choose -f},\tag A3.5$$
$$ {\Cal H}_v(E)={\Cal H}_v^d(E)+{\Cal V}.$$
Here ${\Cal H}_v^d$ (${\Cal V}$) stands for the 
diagonal  (anti-diagonal) part of the operator
${\Cal H}_v$:
$$
{\Cal H}^d_v=
p\left[{d\over dx}-
\pmatrix {\Cal E}&0\cr
0&{\Cal E}^* \cr\endpmatrix\right]+iV\s_3,\quad
{\Cal V}=V\s_2,\quad  \s_2=\pmatrix
0&i\cr
-i&0\cr \endpmatrix,$$
$$ V=V_0+{1\over 2}v,\quad
V_0=- 
{5\over 32}(F+q^\prime)^2 p^{-4}+{1\over 8}q^{\prime\prime} p^{-2}.$$

\par We  consider  ${\Cal H}_v(E)$ 
as an operator in $L_2([R,\infty)\rightarrow {\Bbb C}^2)$
 submitted to the boundary condition:
$z_1(R)+z_2(R)=0.$
We are going to treat the anti-diagonal part ${\Cal V}$ 
as a perturbation. To this purpose, we rewrite (A3.5) in the form
$$z={\Cal A}_vg-{\Cal B}_vg +{\Cal B}_v{\Cal V}z,
\quad
{\Cal A}_v={\Cal H}_v^{d^{-1}},\quad
{\Cal B}_v={\Cal A}_v{\Cal V}{\Cal A}_v,$$
which leads to the following representation 
for the resolvent $R_v$:
$$R_v={\bold p}{\Bbb E}({\Cal A}_vJ -{\Cal B}_vJ
+{\Cal B}_v{\Cal V}\R_v),\tag A3.6$$
where
 ${\bold p}$ is the projection of ${\Bbb C}^2$-vector
onto the first  component: 
${\bold p}=\pmatrix 1&0\cr0&0\cr \endpmatrix$,
the operators $J$, $\R_v(E)$:
$L_2([R,\infty)\rightarrow {\Bbb C})
\rightarrow L_2([R,\infty)\rightarrow {\Bbb C}^2)$
are given by the formulas
$$ Jf={i\over 2}{f\choose -f},\quad
\R_v(E)f={\Bbb E}^{-1}{R_v(E)f\choose {d\over dx} R_v(E)}.$$
Since ${\bold p}{\Bbb E}$, $J$ are bounded in order 
to prove (A3.1) it is sufficient to show that
$\A_v-\A_0$, $\B_v$, $\B_v{\Cal V}\R_v$ are 
compact  in $L_2$. Here $\A_0$ corresponds to $v=0$:
$$\A_0=\left[ p\bigg({d\over dx}-\pmatrix {\Cal E}&0\cr
0&{\Cal E}^* \cr\endpmatrix \bigg)+iV_0\s_3\right]^{-1}.$$

\par Consider the operator $\A_v$. 
It has the form $\A_v=\pmatrix \A^{1}_v&\A^2_v\cr 0&\A^3_v\cr\endpmatrix$,
where $\A^j_v$,  are integral operators with the kernels
$$
\A^1_v(x,y)=a(x,y)\Theta(x-y),\quad \A^3_v(x,y)=-a(y,x)\Theta(y-x),
\quad \A^2_v(x,y)=t(x)t(y),$$
$$a(x,y)={e^{i\int_y^xds(p-Vp^{-1})}\over (p(x)p(y))^{1/2}},
\quad t(x)=
{e^{i\int_R^xds(p-Vp^{-1})}\over (p(x))^{1/2}},\tag A3.7
$$
where $\Theta$ is the Heviside function.
Clearly, for $\im E>0$, one has
$$
|a(x,y)|\leq Cx^{-1/4}y^{-1/4}e^{-\beta\im E(x^{1/2}-y^{1/2})},\quad
x\geq y\geq R, \tag A3.8$$
$$
|t(x)|\leq Cx^{-1/4}
e^{-\beta \im E(x^{1/2}-R^{1/2})},\quad x\geq R,\tag A3.9$$
provided 
$\b<F^{-1/2}$,
$R$ is sufficiently large.
This allows us to conclude that $\A_v$
is  bounded  in $L_2([R,\infty))$ and,
moreover, the component $\A^2_v$ is a Hilbert- Schmidt operator:
$$\|\A^2_v\|_{L_2([R,\infty)^2)}\leq C.\tag A3.10$$
Consider the difference
$a(x,y)- a_0(x,y)$, where
 $a_0(x,y)$ corresponds to $v= 0$:
$$ a_0(x,y)=
{e^{i\int_y^xds(p-V_0p^{-1})}\over (p(x)p(y))^{1/2}}.\tag A3.11$$
Due to (1.2) one has
$$|a(x,y)-a_0(x,y)|\leq |a_0(x,y)|
\left|\int_y^x {v(s)\over p(s,E)}ds\right|$$
$$\leq Cx^{-1/4}y^{-3/4}e^{-\beta\im E(x^{1/2}-y^{1/2})},\quad
x\geq y\geq R, \tag A3.12$$
which implies that
$$\|\A^j_v-\A^j_0\|_{L_2([R,\infty)^2)}\leq C, \quad j=1,3.\tag A3.13$$
As a consequence of (A3.10), (A3.13) we obtain that
the whole difference $\A_v-\A_0$ belongs to the Hilbert- Schmidt class.

\par Consider $\B_v$. 
Its kernel has the form:
$\B_v(x,y)=\C_v(x,y)+\D_v(x,y),$
$$\C_v(x,y)=\pmatrix\C^1_v(x,y) & \C^2_v(x,y)  \cr
                               0&\C^3_v(x,y)\cr\endpmatrix,
\quad \D_v(x,y)=\pmatrix 0 & \D^1_v(x,y)  \cr
                             \D^2_v(x,y)  &0\cr\endpmatrix$$
$$\C^1_v(x,y)=-it(x)t(y)\Psi^1(y),\quad 
\C^2_v(x,y)=-it(x)t(y)\Psi^1(R),$$ 
$$\C^3_v(x,y)=it(x)t(y)\Psi^1(x),$$
$$\D^1_v(x,y)=-ia(x,y)\Psi^2(y),\quad
\D^2_v(x,y)=-ia(x,y)\Psi^1(x),\quad x\geq y\geq R,$$
$$\D^j_v(x,y)=\D^j_v(y,x),\quad j=1,2.$$
Here
$$\Psi^1(x,E)=\int_{x}^\infty dy{e^{2i\int_x^yds(p-Vp^{-1})}\over p(y,E)}
V(y),$$
$$ \Psi^2(x,E)=\int_{R}^x dy{e^{2i\int_y^xds(p-Vp^{-1})}\over p(y,E)}
V(y).$$
Clearly,
$$\|\Psi^i\|_{L_\infty([R,\infty))}\leq C, \quad i=1,2,\tag A3.14$$
which together with (A3.8), (A3.9) implies that
$\C_v$, $D_v$ are  bounded  operators from 
$l^2(L_1)$  to $L_2$:
$$\|\C_v g\|_{L_2([R,\infty))},\,
\|\D_v g\|_{L_2([R,\infty))}
\leq C \|g\|_{l^2(L_1)([R,\infty))},\tag A3.15$$
provided $R$ is sufficiently large. 
Here $l^p(L_q)$ norms are defined by
$$\|f\|_{l^p(L_q)([R,\infty))}=
\left(\sum_{n=0}^\infty \|f\|_{L_q(\triangle_n)}^p
\right)^{{1\over p}},\quad \triangle_n=[n+R,n+1+R].$$
Moreover, since $t(x)$, $t(x)\Psi^1(x)$ belong to
$l^2(L_\infty)$, $\C_v$ is a compact operator. 
The same is true for $\C_v{\Cal V}\R_v$. Indeed,
$\R$ is a bounded operator from
$L_2$  to $l^2(L_\infty)$:
$$\|\R_v g\|_{l^2(L_\infty)}\leq C \|g\|_{L_2},
\tag A3.16$$
see [16], for example. As a consequence, $\C_v{\Cal V}\R_v$
is compact in $L_2$.
\par We next focus on the contribution of $\D_v$ in (A3.6).
Consider the functions $\Psi^j(x,E)$, $j=1,2$.
They admit the representations
$$\Psi^j(x,E)=\Psi^j_0(x,E)+O(x^{-1/2}),\tag A3.17$$
where
$$\Psi^1_0(x,E)=\int_{x}^\infty dy{e^{2i\int_x^yds(p_0+Ep_0^{-1})}
\over p_0(y)}
v(y),$$
$$\Psi^2_0(x,E)=\int^{x}_R dy{e^{-2i\int_x^yds(p_0+Ep_0^{-1})}
\over p_0(y)}
v(y),$$
 $p_0(x)=p(x,0)$.
It is not difficult to show that
for $v\in L_1({\Bbb T})$,
$$|\Psi^1_0(x,E)|\leq Cx^{-1/4}, \tag A3.18$$
see for example, [16], appendix 3.
Due to the representation
$$\Psi^2_0(x,E)
=-\overline{\Psi^1_0(x,E)}+e^{2i\int^x_Rds(p_0+Ep_0^{-1})}
\overline{\Psi^1_0(E,R)}
$$
$$+2\im E
\int_R^x dy{e^{2i\int^x_yds(p_0+Ep_0^{-1})}
\over p_0(y,E)}\Psi^1_0(y,E),$$
the same inequality is valid for $\Psi^2_0(x,E)$:
$$|\Psi^2_0(x,E)|\leq Cx^{-1/4}.$$
As a consequence,
one gets the following estimate for $\D_v$:
$$\|\D_vg\|_{L_2([A,\infty))}\leq CA^{-1/4}\|g\|_{l^2(L_1)([R,\infty))},
\tag A3.19$$
for any $A$, $A\geq R$.
\par It follows directly from the definition of $\D_v$ that
$\D_vg\in W^{1,1}_{loc}$, provided $g\in l^2(L_1)$ and moreover,
one has the estimate
$$\|x^{-1/2}{d\over dx}\D_vg\|_{l^2(L_1)}
\leq C\|g\|_{l^2(L_1)}.\tag A3.20$$
(A3.16), (A3.19), (A3.20) 
imply that both $\D_v$ and $\D_v{\Cal V}\R_v$
are compact operators in $L_2$.\hfill $ \blacksquare$

\subhead Acknowledgment
\endsubhead
It is a pleasure to thank V.S. Buslaev,
J. Sjostrand 
for  helpful discussions.

\vskip10pt
\subhead
 References
\endsubhead
\vskip4pt
\item{1.} Ao, P.: Absence of localization in 
energy space of a Bloch electron driven
by a constant electric force,  Phys Rev. B  {\bf 41}, 3998-4001 (1990).
\vskip4pt
\item{ 2.} Asch, J., Duclos, P. and Exner, P.:  
Stark-Wannier Hamiltonians
with pure point spectrum, in 
 {\it Differential equations, asymptotic analysis, and mathematical
physics}, Potsdam 1996, pp. 10--25, Math. Res. {\bf 100}, 
Akademie Verlag, Berlin, 1997. 
\vskip4pt
\item{ 3.} Asch, J., Duclos, P. and Exner, P.:  
Stability of driven systems
with growing gaps, quantum rings, and Wannier ladders,
 J. Statist. Phys. {\bf 92}, 
1053--1070 (1998).
\vskip4pt
\item{4.}  Asch, J., Duclos, P., Bentosela, F. and Nenciu, G.:
On the dynamics of crystal electrons, high momentum regime,
 J. Math. Anal. Appl. {\bf 256}, 99--114
(2001).
\vskip4pt
\item{5.}
Bentosela, F.,  Carmona, R., Duclos, P., Simon, B.,
 Souillard, B. and  Weder, R.:
Schr\"odinger operators with an electric field 
and random or deterministic potentials,
 Comm. Math. Phys. {\bf 88}, 387--397 (1983).
\vskip4pt
\item{6.} 
Bourgain, J., Schlag, W.: 
Anderson localization for Schr\"odinger operators
on ${\Bbb Z}$ with strongly mixing potentials, 
Comm. Math. Phys. {\bf 215},
143--175 (2000).
\vskip4pt 
\item{7.} Buslaev, V.S.: Kronig-Penney electron 
in homogeneous electric field,
Amer. Math. Soc. Transl. Ser.2, 189, 
Amer. Math. Soc., Providence, RI, 1999.
\vskip4pt
\item{8.}
 Christ, M. and Kiselev, A.:
Absolutely continuous spectrum  of Stark operators, Ark. Mat. {\bf 41},
no. 1, 1-33 (2003).
\vskip4pt
\item{9.} Delyon, F., Simon, B. and Souillard, B.:
From power pure point to continuous spectrum
in disordered systems, Ann. Inst. H. Poincar\'e Phys. Th\'eor. {\bf 42}, 
no. 3, 283--309 (1985). 
\vskip4pt
\item{10.} Exner, P.: The absence of the absolutely continuous
spectrum for $\delta^\prime$ Wannier-Stark ladders,  J. Math. Phys,
{\bf 36(9)}, 4561-4570 (1995).
\vskip4pt
\item{11.} Figotin, A., Pastur, L.: 
Spectra of random and almost periodic operators,
Springer-Verlag, Berlin, 1992.
\vskip4pt
\item{12.} Gilbert, D.J, Pearson, D.B: On subordinacy and analysis of 
the spectrum of one-dimensional Schr\"odinger operators,
J. Math. Anal. and Appl. {\bf 128}, 30-56 (1987).
\vskip4pt
\item{13.} Gilbert, D.J: On subordinacy and analysis of 
the spectrum of Schr\"odinger operators with two singular endpoints,
Proc. Royal Soc.  Edinburgh, {\bf 112A}, 213-229 (1989).
\vskip4pt
\item{14.} Kiselev, A., Last, Y. and  Simon, B.:
Modified Pr\"ufer and
EFGP transforms and the spectral analysis of one-dimensional 
Schr\"odinger operators,
 Comm. Math. Phys. {\bf 194}, 1-45  (1998). 
\vskip4pt
\item{15.} Maioli, M. and A.Sacchetti, A.: Absence of the absolutely
continuous spectrum for Stark-Bloch operators
with strongly periodic potentials, J. Phys. A {\bf 28}, 1101-1106 (1995).
\vskip4pt
\item{16.} Perelman, G.: On the absolutely continuous spectrum
of Stark operators, Comm. Math. Phys. {\bf 234}, 359-381 (2003).
\vskip4pt
\item{17.} Pozharski, A. A.: On operators of Wannier-Stark type 
with singular potentials, St. Petersburg Math. J. {\bf 14}, 
no. 1, 119--145 (2003). 
\vskip4pt
\item{18.} Weidmann, J.: 
{\it Spectral Theory of Ordinary Differential Operators}.
LNM 1258, Springer-Verlag, Berlin, 1987.

\end